%Paper: hep-th/9306072
%From: physics@IREARN.BITNET
%Date: Wed, 16 Jun 1993 16:16:53 -0305
%Date (revised): Mon, 21 Jun 1993 14:09:49 -0305

\magnification=1200
\def\caz{{\cal Z}}
\rightline {SUTDP/11/93/72}
\rightline {May, 1993}
\bigskip \bigskip
\centerline {{\bf A UNIFIED SCHEME FOR MODULAR INVARIANT }}
\bigskip
\centerline {{\bf PARTITION FUNCTIONS OF WZW MODELS}}
\bigskip \bigskip \bigskip
\vskip 1.0 cm
\centerline  { M. R. Abolhassan$i^a$,\ \ F. Ardala$n^{a,b}$}
\vskip 0.3 cm

\centerline { \it $ ^a$Department of Physics, Sharif University of
 Technology}
\centerline {P.O.Box 11365-9161, Tehran, Iran}
\centerline {\it $^b$Institute for studies in Theoretical Physics and
Mathematics}
\centerline {P.O.Box 19395-1795, Tehran, Iran}

\bigskip \bigskip \bigskip
\vskip 2.0 cm
\centerline{\bf Abstract}
We introuduce a unified method which can be applied to any WZW
model at arbitrary level to search systematically for modular invariant
physical partition functions. Our method  is based essentially on modding
out a known theory on group manifold $G$ by a discrete group $\Gamma$.
 We apply our method to $\widehat {su(n)}$  with $n=2,3,4,5,6$,
 and to $\widehat {g_2}$ models, and obtain all the known partition functions
 and some new ones, and give explicit expressions for all of them.

\vfill \eject
%\pageno=1
\noindent {\bf 1.Introduction}

\noindent Conformal field theories (CFT's) play an important role in
two dimensional critical statistical mechanics and string theory , and
 has been extensively studied in the past decade.$^{1-4}$
 Among these theories WZW models have attracted considerable attentions,
 because as two dimensional rational conformal field theories they are
 exactly solvable,$^{5,6}$ and most known CFT's can be obtained from them via
 coset  construction.$^7$ Moreover these models explicitly appear in
 some statistical models like quantum chains, and describe their critical
 behavier.$^8$
 WZW models in addition to conformal symmetry, have an infinite dimensional
symmetry whose currents satisfy a Kac-Moody algebra $\hat g$ at some level
$k$. The partition function of a WZW model takes the form
$$Z(\tau,\bar\tau)=\sum \chi_\lambda(\tau)\,M_{\lambda,{\lambda'}}\,
\chi_{\lambda'}^*(\bar\tau),\eqno(1.1)$$

\noindent where $\chi_\lambda$ is the character of the affine module
whose highest weigth (HW) is $\lambda$,
$M_{\lambda,\lambda '}$  are positive integers which determine how many times
the HW representations $\lambda$, $\lambda'$ in the left and right moving
sectors couple with each other, and the sum is over the finite set of
integrable representations (see Subsec. 2.1. for a precise definition).
 For the consistency of a physical theory it is necessary that the partition
 function (1.1) be invariant under the modular group of the torus.$^9$
Construction and classification of partition functions of WZW models
has been the goal of a large body of work in the past few years. However, up
to now only the classifications of
$\widehat {su(2)}$,$^{10}$ and $\widehat {su(3)}$$^{11-13}$ at arbitrary
level,
and of simple affine Lie algebras at level one$^{14,15}$ have been completed.

 To find moudular invariant partition functions of WZW models a number of
 methods have been used: Automorphism of Kac-Moody algebras,$^{16,17}$ simple
 currents,$^{18,19}$ conformal embedding,$^{20-22}$ automorphism of
 the fusion rules of the extended chiral algebra,$^{23,24}$ lattice
 method,$^{26,27}$ and finally
 direct computer calculations.$^{28}$ In these efforts many modular invariant
partition functions have been found, and may be arranged in three broad
categories:

 i)\ \  Diagonal Series  $-$ For every WZW model with a simply
connected group manifold, there exists a physical modular invariant theory
 with diagonal matrix $M_{\lambda ,\lambda '}=\delta _{\lambda ,\lambda '}$.
$^{9,17}$ They are often designated as a member of the  $A$ series.

 ii)\ \  Complementary Series $-$ There are some nondiagonal series
 for every WZW model whose Kac-Moody algebra has a nontrivial
centre,$^{16,17}$ associated to subgroups of the centre.
They are often designated as members of the $D$ series.

 iii)\ \  Exceptional Series $-$ In addition to the above two series,
 WZW models have a number of nondiagonal partition functions which occur
 only at certain levels. They are called $E$ series.

Some of the known $E$ series have been found by the conformal embedding
method (see e.g. Ref. 13), some by utilizing the nontrivial automorphism
of the fusion rules of the extended algebra,$^{24,25}$ and some others
by computer calculations.$^{28}$
Alhough many exceptional partition functions have been obtained by these
methods, however they don't follow from a unified method and prove to be
 impractical for high rank groups and high levels. Furthermore, these
methods don't answer the question of why there are exceptional partition
functions  only at certain levels.
It must be mentioned that corresponding to any physical theory, there exists
a charge conjugation, c.c., counterpart, such that
$M_{\lambda,\lambda'}^{(c.c.)}=M_{\lambda,C(\lambda')}$, where
$\scriptstyle C(\lambda)$ is the complex conjugate representation of
$\lambda$.

In this paper a unified approach which we call {\it orbifold-like} method,
 is presented and shown to lead to all the known nondiagonal theroies.
 The method is easily applied to high rank groups at arbitrary level.
 The organization of the paper is as follows:
In Sec. 2, we briefly review some characteristic features of Kac-Moody
algebra, such as their unitary highest weight representations and
modular transformation properties of their characters.
Then we present our approach. We start with some known theory whose partition
 function is $Z(G)$ and mod it out by a discrete group $\Gamma$.
 We will take $\Gamma$ to be a cyclic group $\caz_N$. It
is not  necessarily a subgroup of the centre of $\hat g$. In order for
the  modding to gives rise to a modular invariant combination with rational
 coefficients, certain relations must be satisfied, which will be explicated.
In Sec. 3, we apply our method to $\widehat {su(n)}$ models with
$n=2,3,4,5,6$ and as an example of a non-simply laced affine Lie algebra, to
$\widehat {g_2}$ models, and generate all the known nondiagonal theories
and some new exceptional ones.
In Sec. 4, we conclude with some remarks.
In Appendix $A$, we gather some formulas and relations that are used in the
body of the paper.

\

\noindent {\bf 2. The Orbifold-like method}

\noindent Observing that all modular invariant partition functions of
$\widehat {su(n)}$ models at level one are obtained by modding out the group
 $SU(n)$ by subgroups of its centre $^{15}$, the authors in Ref. 29
 for the first time found that not only the $D$ series of $\widehat {su(2)}$
 WZW models are obtained with modding out the diagonal theories by
 the $\caz_2$ centre, but also all their
 exceptional partition functions can be found by modding out the $A$ or $D$
 series by a $\caz_3$ which is not obviously a subgroup of the centre.

In this paper we are going to generalize that work to WZW models with
any associated affine Lie algebra at arbitrary level. We have called our
approach {\it orbifold-like} method, because the finite group which one uses
in modding may not be the symmetry of a target manifold.
 Befor going to the details of our approach, we collect in the following
 subsection, some facts about untwisted Kac-Moody algebras and set up our
 notations.

\

\noindent {\bf 2.1. Preliminaris and Notations}

\noindent A WZW model is denoted by $\hat g_k$  where its affine symmetry
algebra is the untwisted Kac--Moody algebra $\hat g$  associated with
a compact Lie algebra $g$, and the positive integer number $k$ is the level
of the Kac-Moody algebra (see Refs. 30,31 for details).
The primary fields of the model can be labelled by the highest weight
representations of horizontal Lie algebra $g$. In the basis of fundamental
weights $\omega_i$ of g these HW representations are expressed by
$\lambda = \Sigma_{i=1}^r\lambda_i \omega_i$
where $\lambda_i$'s are positive integers (Dynkin labels) and $r$ is the rank
of $g$. Imposing unitarity condition, restricts the numbre of
HW representations which appear in a theory at a given level.
These representations which are usually called integrable representations
satisfy the relation ${2\over\Psi^2} (\lambda,\Psi)\leq k$,
where $\Psi$ is the highest root of $g$. The set of integrable
representations is called the fundamental domain and is denoted by $B_h$,
where the height $h$ is defined by $h=\check h+k$ and $\check h$ is the
dual Coxeter number of $g$. We choose the normalization such that
$\Psi^2 = 2$.
The characters of integrable representations at a given level $k$ constitute
a linear unitary representation of modular group of the torus (see e.g.
Ref. 32). The character of an integrable HW representation transforms
under the action of the generators of the modular group
$S:(\tau \to -1/\tau )$ and $T:(\tau \to \tau +1)$ as
$$\eqalign{\chi_\lambda(-1/\tau)&=C\>\sum_{\lbrace\lambda'\in {B_h}\rbrace}
\sum_{\lbrace\omega\in {W(G)}\rbrace}\varepsilon (\omega)
e^{({2\pi i\over h})\left(\tilde\lambda,\omega(\tilde\lambda')\right)}
\quad\chi_{\lambda'}(\tau)\cr
\chi_\lambda(\tau +1)&=e^{\pi i\bigl(\tilde\lambda^2/ h
-\rho^2/\check h\bigr)}\quad\chi_\lambda(\tau),\cr}\eqno(2.1)$$
\noindent where $C={i^{\vert\Delta_+\vert}\over(k+\check h)^{r/2}}
\left({voll.\,cell\,of\,Q^* \over voll.\,cell\,of\,Q}\right)^{1/2}$,
$Q$ is the coroot lattice, $Q^*$ its dual lattice, and $\tau$,
 $\Delta_+$, W , $\epsilon(\omega)$ are the parameter of the torus,
 the number of positive roots, the Weyl group, and the determinant
 of the Weyl reflection $\omega$, respectively.
In formula (2.1) $\tilde \lambda =\lambda+\rho$ and $\rho$ is the sum of
the fundamental weights. It is more easier to work with $\tilde \lambda$
than $\lambda$, but hereafter we omit the tilde sign of
$\tilde \lambda$.
Thus, wherever we refer to the fundamental domain, we mean:
$$B_h = \biggl\lbrace \lambda=\sum_{i=1}^r m_i \omega_i\; \Big\vert\;
\sum_{i=1}^r m_i \check h_i < {h = k + \check h}\biggr\rbrace,\eqno(2.2)$$

\noindent where $\check h_i$'s are the dual Dynkin labels and appear in the
expansion  of $\Psi$ in terms of $\alpha_i$'s, the simple roots of $g$:
$\Psi/(\Psi)^2=\sum_{i=1}^r \check h_i \alpha_i/(\alpha_i)^2$,
and $\check h=\sum_{i=1}^r\check h_i$.

\

\noindent  {\bf 2.2. Orbifold approach}

\noindent  It was noted in Ref. 35 that each complementary series
obtained in Ref. 17 is actually the partition function of an orbifold,
 which is costructed via modding the covering group $\tilde G$ by a subgroup
 of its centre. For the construction  of this partition function,
one starts with the WZW  theory defined on  the group manifold $\tilde G$
and impose boundary conditions on the fields up to the action of some
subgroup $\Gamma$ of the centre of $\tilde G$:
$$\Phi(\sigma_1+2\pi,\sigma_2)=h_1\Phi(\sigma_1,\sigma_2)
\quad;\quad\Phi(\sigma_1,\sigma_2+2\pi)=h_2\Phi(\sigma_1,\sigma_2),
\eqno(2.3)$$
\noindent where $\sigma_1 ,\sigma_2 $ are coordinates on the torus and
$h_1,h_2$
are some elements of $\Gamma$. Let us denote by $(h_1,h_2)$ the contribution
 to  the partition function from the twisted sector with boundary conditions
 (2.3). In order for the theory $G/\Gamma$ to be modular invariant, all
 twisted sectors must be included.$^{9,34}$ So the partition function
 of the new theory can be written in the following form
$$Z( G/\Gamma)={1\over \vert\Gamma\vert} \sum_{\scriptstyle h_1,h_2 \in
\Gamma \atop \scriptstyle [h_1,h_2]=0} (h_1,h_2),\eqno(2.4)$$
\noindent where $\vert\Gamma\vert$ is the order of $\Gamma$.
For $\Gamma=\caz_N$ which is the interesting group for us, eq. (2.4)
reduces to:
$$Z({ G/\caz_N})={1\over N} \sum_{\alpha,\beta =1}^N
(h^{\alpha},h^{\beta}).\eqno(2.5)$$
It can easily be seen that using the untwisted sector, defined by
$$Z_1 ({ G/\caz_N})={1\over N} \sum_{\alpha=1}^N (1,h^{\alpha}),\eqno(2.6)$$
and acting properly on it by the generators  of the modular group, $S$ and
$T$, one can obtain  the full partition function $Z({ G/\caz_N})$.$^{9}$
In Ref. 35 the following formula is drived for $Z({ G/\caz_N})$
when ${\caz}_N$ is a subgroup of the centre of $ G$ and $N$ is prime:
$$Z({G/\caz_N})=\biggl[\sum_{\alpha=1}^N\Bigl(T^{\alpha} S+1\Bigr)Z_1
({G/\caz_N})\biggr]-Z(G). \eqno(2.7)$$

Concerning the method mentioned above we make two crucial observations.
First, the orbifold method can be similarly applied to the case where $G$
is not the covering group, and even when $\caz_N$ is not a subgroup of
the centre of $G$, but is a symmetry of the classical theory.
However, in those cases in order for the modding
to be meaningful, i.e., the sum of the terms in the {\it bracket} of eq.
 (2.7) and therefor the whole expression be modular invariant with real
coefficients, the following relation must be satisfied
$$T^N SZ_1({G/\caz_N})=SZ_1({G/\caz_N}).\eqno(2.8)$$
Secondly, for the case when $N$ is not prime, the expression (2.7) does not
completely describe an orbifold partition function and in order to generate
the full partition function $Z(G/\caz_N)$, some additional terms must be
included into the {\it bracket}. But then, extra constraints beyond (2.8)
have to be satisfied for modular ivariance (see eq. (A.3)). We call the
appropriate moddings for a given theory which satisfy these constraints,
{\it allowed} moddings. We have collected in Appendix $A$, a list of
formulas for the cases of interest to us.

Our strategy in finding the nondiagonal WZW theories with the affine symmetry
algegra $\hat g_k$ is as follows. First we start with a diagonal theory
and mod it out by some group $\caz_N$. The untwisted part
of the partition function $Z_1$, is realized by representing the action
of $\caz_N$ on the characters of HW representations in the left$-$moving
sector as
$$p\cdot\chi_\lambda (\tau)=e^{{2\pi i\over N }\beta(\lambda^2-\rho^2)} \;
\chi_\lambda(\tau)\eqno(2.9)$$
\noindent where $p$ is the generator of $\caz_N$, $\beta$ is the smallest
 rational number such that $\beta (\lambda^2-\rho^2)$ is an integer.
Therefor the the untwisted part $Z_1$ consists of left$-$moving
representations with $\beta(\lambda^2-\rho^2)=0 mod\,N$.
It must be mentioned that the realization of $\caz_N$ in (2.9) is such that
it gives the exact form of $D_h$ series for $\widehat {su(2)}$ models,
then we have simply generalized it to account for all nondiagonal
partition functions. Now we act on the untwisted part $Z_1$ by the
operators $S$ and $T$ according to the {\it bracket} in the r.h.s of
the corresponding formula of $Z({G/ \caz_N})$, and check in every step
if the related condition (A.3) is satisfied, and finally we calculate the
sum. Then we encounter three cases:

{\bf Case\ I)}\  All the terms $\chi_\lambda\bar\chi_{\lambda'}$ which
appear in the sum have positive rational coefficients. This indicates that
the sum is a positive linear combination of modular invariant partition
functions.
Sometimes after subtracting some previously known partition functions from
this, a new partition function will appear. See for example, the case of
modding  $D_8^{(2)}$ by $\caz_{16}$ in $\widehat {su(4)}$ models on
page 19.

{\bf Case\ II)}\ Some of the terms in the sum have negative coefficients.
However after adding or subtracting some known partition functions
or/and some modular invariant combinations of characters that have been
found in the process of the previous moddings, a new modular invariant
partition function will be found. See for example, the case of modding
$D_8$ by $\caz_9$ in $\widehat {su(3)}$ models on page 15.

{\bf Case\ III)}\ Some of the terms in the {\it bracket} have negative
coefficients but after subtracting from it some known physical partitions
or/and some modular invariant combinations of characters,
at most a new modular invariant combination will be obtained, which is
not a partition function. See for example, the case of modding
$D_{24}$ by $\caz_2$ in $\widehat {su(3)}$ models on page 16.

\

\noindent {{\bf 3.Applications}}

\noindent In this section we apply our method to $\widehat {su(n)}$ WZW
models with $n=2,3,4,5,6$, and as an example of a nonsimply-laced affine
Lie algebra to $\widehat {g_2}$ WZW models.

\

\noindent {\bf 3.1. $\widehat {su(2)}$  WZW models}

 For $\widehat {su(2)}$  models besides the usual diagonal series
$A_h=\sum \nolimits_{\lbrace\lambda\in {B_h}\rbrace} \vert \chi_\lambda
\vert ^2$ at each level, where
$$B_h = \Bigl\lbrace \lambda=m\,\omega\; \big \vert \; 1 \leq m < h=k+2
\Bigr\rbrace\eqno(3.1)$$
is the fundamental domain and $\omega$ is the fundamental weight of $su(2)$,
 and a nondiagonal $D$ series at even levels, three
 exceptional modular invariant partition functions  have been found at
 levels $k =10,16,28$; and it has been shown that this set completes
 the classification of $\widehat {su(2)}$ WZW models.$^{10}$
In what follows we will review the results of Ref. 29, where it was
shown that the exceptional partition functions can be obtained by our
orbifold method, and present some further calculations. The action of
the $\caz_N$ group on the characters of the left$-$moving HW representations
of $su(2)$ is defined due to eq. (2.9) by
$$p \;\cdot\; \chi_m=e^{{2\pi i\over N}(m^2-1)}\;\chi_m,\eqno(3.2)$$
\noindent where p is the generator of $\caz_N$.
Hereafter the HW representation $\lambda$ is designated by its Dynkin
label $m$. Thus the untwisted part of a partition function, consists of HW
 representations $\lambda=m\,\omega$ with $m^2=1\, mod\,N$.

\

\noindent {\bf D - Series}

These modular invariant partition functions are obtained by modding out the
diagonal series by $\caz_2$, the centre of $SU(2)$.
The untwisted part of the partition function $Z_1$ is given by
$$Z_1(A_h/{\caz_2}) = \sum\nolimits_{\lbrace \lambda\, \vert\, m \, odd
\rbrace}\vert \chi_\lambda \vert ^2.\eqno(3.3)$$
The partition function $Z( G/\caz_2)$ can be calculated using eq. (2.7)
with $N=2$. The calculation is straightforward. First the untwisted
part (3.3) is written in the following form
$$Z_1(A_h/\caz_2) ={1\over{\vert W \vert}} \sum\nolimits_{\lbrace \lambda
\in  W B_h \, ;\;  m\, odd \rbrace} \vert
\chi_\lambda \vert ^2,\eqno(3.4)$$
using the identity
$\chi_{\omega(\lambda)}=\epsilon (\omega) \chi_{\lambda}$, where $WB_h$ is
the Weyl reflection of the fundamental domain $B_h$.
Then we rewrite (3.4) in the following form
$$Z_1(A_h/\caz_2) ={1\over{\vert W \vert}} \sum\nolimits_{\lbrace \lambda
\in (Q^*/h\,Q) \,;\, m \,odd \rbrace} \vert \chi_\lambda \vert ^2,
\eqno(3.5)$$
noting that $\chi_0=\chi_h=0$. On the other hand the lattice
$Q^*\over h \, Q$ consists of the following HW representations:
$${Q^*\over h \, Q} =\Bigl\lbrace\lambda=m\,\omega\;\Bigm\vert\; 1\leq m
\leq 2h\Bigr\rbrace.\eqno(3.6)$$
Now acting by the operator $S$ in (2.1) on eq. (3.5) we obtain:
$$\eqalign{SZ_1&={1\over 2h}{1\over2!} \sum_{\scriptstyle\lambda
\in Q^*/hQ\atop
\scriptstyle m \, odd}
\sum_{\scriptstyle \lambda',\lambda''\in B_h\atop \scriptstyle\omega',
\omega''\in W}
\!\varepsilon(\omega'\omega'') \exp\biggl[{2\pi i\over h}\Bigl(\lambda,\omega
'(\lambda')-\omega''(\lambda'')\Bigr)\biggr]
\chi_{\lambda'} \bar\chi_{\lambda''}.\cr}\eqno(3.7)$$
The sum over $\lambda$ is easily done using eq. (3.6), and it appears
that the sum over one of the two Weyl groups can be factored out and give
an overall factor equal to the order of Weyl group. Finally the sum over
the other Weyl group must be done. Substituting (3.7) in eq. (2.7), we
easily do the sum and find a nondiagonal partition function at every
even level given by
$$D_h\equiv Z(A_h/\caz_2) =\sum_{m\,odd=1}^{h-1}\big\vert\chi_m\big\vert^2+
\sum_{m\,odd=1}^{h/2-2}(\chi_m \bar\chi_{h-m}+c.c.)+2
\big\vert\chi_{h/2}\big\vert ^2\eqno(3.8a)$$
for $h=2\,mod\,4$, and
$$D_h\equiv Z(A_h/\caz_2) =\sum_{m\,odd=1}^{h-1}\big\vert\chi_m\big\vert ^2+
\sum_{m\,even=2}^{h/2-2}(\chi_m \bar\chi_{h-m}+c.c.)+
\big\vert\chi_{h/2}\big\vert ^2\eqno(3.8b)$$
for $h=0\,mod\,4$.
these are exactly the $D$ series given in Ref. 10.

\

\noindent {\bf E - Series}

One expects to find possibly, exceptional partition functions at levels
$k = 4,10,28 $ according to the following conformal embeddings:$^{20}$
$$\widehat {su(2)}_{k=4} \subset \widehat {su(3)}_{k=1},\quad
\widehat {su(2)}_{k=10} \subset \widehat {(B_2)}_{k=1},\quad \widehat
 {su(2)}_{k=28} \subset \widehat {(g_2)}_{k=1}.\eqno(3.9)$$

{\bf 1.}\ At level $k = 4\;(h=6)$, we start with $A_6$ and mod it out by
$\caz_3$, and check that the eq. (2.8) is satisfied; then we do the sum
in the {\it bracket} of eq. (2.7) with $N=3$, and  finally get
$$\Bigl[\sum_{\alpha=1}^3 T^\alpha SZ_1+Z_1\Bigr]=4A_6-D_6.\eqno(3.10)$$
We continue modding by {\it allowed} $\caz_N$'s up to $N=48$, but nothing
more than $A_6$ and $D_6$ appears. For example, in the case $N=6$ we
obtain
$$\Bigl[\bigl(\sum_{\alpha=1}^6 T^\alpha+\sum_{\alpha=1}^3T^\alpha ST^2
+\sum_{\alpha=1}^2T^\alpha ST^3\bigr)SZ_1+Z_1\Bigr] =4A_6+D_6.\eqno(3.11)$$
Then we start with $D_6$, mod it out by {\it allowed} moddings but also
nothing more is found. For example, in modding by $\caz_3$ we find
$$\Bigl[\sum_{\alpha=1}^3 T^\alpha SZ_1+Z_1\Bigr]=2D_6.\eqno(3.12)$$
This is not surprising, since it can easily be seen that $D_6$ given by
$$D_6=\big\vert\chi_1+\chi_5\big\vert^2
+2\big\vert\chi_3\big\vert^2,\eqno(3.13)$$
exactly corresponding to conformal embedding
$\widehat {su(2)}_{k=4} \subset \widehat {su(3)}_{k=1}.$

{\bf 2.}\  At level $k=10\; (h=12)$, we start with $A_{12}$ and mod it out
by $\caz_6$ which is allowed. After doing the sum in the {\it bracket} of
eq. (A.5), we encounter the case I of Subsec. 2.2., which after subtracting
the known partition functions $A_{12}$ and $D_{12}$ each with mutiplicity
$2$, we obtain the exceptional partition function $E_6$, which in our
notation is described by $E_{12}$
$$\Bigl[\bigl(\sum_{\alpha=1}^6 T^\alpha+\sum_{\alpha=1}^3T^\alpha ST^2
+\sum_{\alpha=1}^2 T^\alpha ST^3\bigr)SZ_1+Z_1\Bigr] =2A_{12}
+2D_{12}+2E_{12},\eqno(3.14)$$
\noindent where
$$E_{12}=\big\vert\chi_1+\chi_7\big\vert^2+\big\vert\chi_4+\chi_8\big\vert^2
+\big\vert\chi_5+\chi_{11}\big\vert^2.\eqno(3.15)$$
In Ref. 29, $E_{12}$ was obtained by modding $A_{12}$ or $D_{12}$ by
$\caz_3$, but there, we encounter the case II of Subsec. 2.2. We also
get $E_{12}$ in modding $D_{12}$ by $\caz_6$; the result is exactly the
same as eq. (3.14).

{\bf 3.}\  At level $k=16\; (h=18)$, we start with $A_{18}$ and mod it out
by $\caz_6$ which is allowed. After doing the sum  in the {\it bracket} of
eq. (A.5), we encounter the case I of Subsec. 2.2., which after subtraction
$A_{18}$ of mutiplicity $3$, the exceptional partition function $E_{18}$
is obtained:
$$\Bigl[\bigl(\sum_{\alpha=1}^6 T^\alpha+\sum_{\alpha=1}^3T^\alpha ST^2
+\sum_{\alpha=1}^2T^\alpha ST^3\bigr)SZ_1+Z_1\Bigr] =3A_{18}+3E_{18},
\eqno(3.16)$$
\noindent where
$$E_{18}=\big\vert\chi_1+\chi_{17}\big\vert^2+\big\vert\chi_5+\chi_{13}
\big\vert^2+\big\vert\chi_7+\chi_{11}\big\vert^2+\big\vert\chi_9\big\vert^2
+\bigl( \chi_9(\overline{\chi_3+\chi_{15}})+c.c.\bigr).\eqno(3.17)$$
We also find $E_{18}$  in modding $D_{18}$ by the {\it allowed} modding like
$\caz_3$:
$$\Bigl[\sum_{\alpha=1}^3 T^\alpha SZ_1+Z_1\Bigr] =D_{18}+2E_{18}.
\eqno(3.18)$$

{\bf 4.}\  At level $k=28\; (h=30)$, we start with $D_{30}$ and mod it out by
$\caz_3$ which is allowed. After doing the sum  in the {\it bracket} of
eq. (2.7), we encounter the case I of Subsec. 2.2, which after subtraction
 $D_{12}$ of mutiplicity $2$, leads to the exceptional partition function
$E_{30}$:
$$\Bigl[\sum_{\alpha=1}^3 T^\alpha SZ_1+Z_1\Bigr]=2D_{30}+E_{30},
\eqno(3.19)$$
\noindent where
$$E_{30}=\big\vert\chi_1+\chi_{11}+\chi_{19}+\chi_{29}\big\vert^2+
\big\vert\chi_7+\chi_{13}+\chi_{17}+\chi_{23}\big\vert^2.\eqno(3.20)$$

So in this subsection we have generated, by orbifold method, not only the
partition functions which correspond to a conformal embedding like $E_{12}$
and $E_{30}$,$^{22}$ but also the one which follows from a nontrivial
automorphism of the fusion rules of the extended algebra i.e.
 $E_{18}$.$^{24}$
It is interesting to notice that all the nondiagonal partition functions
of $su(2)$ are obtained from moddings  by $\caz_N$'s, with $N$ a divisor of
$2h$ and $h=k+2$.

\

\noindent {\bf 3.2. $\widehat {su(3)}$ WZW models}

\noindent For $\widehat {su(3)}$  models besides the usual diagonal series
$A_h=\sum \nolimits_{\lbrace\lambda\in {B_h}\rbrace} \vert \chi_\lambda
\vert ^2$ at each level, where
$$B_h = \Bigl\lbrace \lambda\; \big\vert
\; 2\leq \sum_{i=1}^2 m_i < h = k + 3\Bigr\rbrace\eqno(3.21)$$
and $\lambda=\Sigma_{i=1}^2m_i\,\omega_i$, and a nondiagonal $D_h$ series
at each level; four exceptional modular invariant partition functions have
been found at levels $k = 5, 9, 21$.$^{12,13}$ Recently, it was shown
that this set completes the classification of $\widehat {su(3)}$ WZW models.
$^{11}$ In what follows we obtain all of these by the orbifold method. We
define the action of the $\caz_N$ group on the characters of the
left$-$moving HW representations of $su(3)$ by
$$p \;\cdot\; \chi_{(m_1,m_2)}=e^{{2\pi i\over N}(m_1^2+m_2^2+m_1m_2-1)}
\;\chi_{(m_1,m_2)}\eqno(3.22)$$
\noindent where p is the generator of $\caz_N$.
Thus the untwisted part of a partition function, consists of left$-$moving
 HW representations  which satisfy: $m_1^2+m_2^2+m_1m_2=1\, mod\,N$.

\

\noindent {\bf D- Series}

These modular invariant partition functions are obtained by modding out the
diagonal series $A_h$ by $\caz_3$, the centre of $SU(3)$.
The untwisted part of the partition function $Z_1$ is given by
$$Z_1(A_h/{\caz_3}) ={ \sum\nolimits_{\lbrace \lambda \vert m_1 - m_2 =
0\,mod\,3\rbrace} \vert \chi_\lambda \vert ^2}.\eqno(3.23)$$
The partition function $Z( G/\caz_3)$ can be calculated using eq. (2.7) with
$N=3$. Following the same recipe mentioned in section 3.1, first we write
the untwisted part (3.23) in the following form
$$Z_1(A_h/\caz_3) ={1\over{\vert W \vert}} \sum\nolimits_{\lbrace \lambda
\in  W B_h \vert m_1 - m_2=0 mod3 \rbrace} \big\vert
\chi_\lambda \big\vert ^2.\eqno(3.24)$$
It is more convenient to write HW representations in the basis
consisting of the simple root $\alpha_1 $ and the corresponding fundamental
weight $\omega_1 $. In this basis
$${Q^*\over h \, Q} =\Bigl\lbrace\lambda=m\,\omega_1+m'\alpha_1\;\big\vert\;
1\leq m \leq 3h+2 \, ;\, -1\leq m'\leq h-2\Bigr\rbrace.\eqno(3.25)$$
so that, just as in eq. (3.5) we can rewrite eq. (3.24) in the form,
$$Z_1(A_h/\caz_3) ={1\over 3!} \sum\nolimits_{\lbrace \lambda = m \omega_1
+m'\alpha_1 \vert 3\leq m \leq 3h+2,\;-1 \leq m' \leq h-2 ; \; m=0 mod \,
3\rbrace} \big\vert \chi_\lambda\big\vert ^2.\eqno(3.26)$$
Now the action of the operator $S$ in eq. (2.1) on eq. (3.26) can be
calculated as mentioned in Subsec. 3.1.
Substuting the untwisted part (3.26) in eq. (2.7) with $N=3$, and doing the
sum in the {\it bracket}, finally we obtain at each level a nondiagonal
partition function denoted by $D_h$:
$$\eqalignno {D_h \equiv Z(A_h/\caz_3)&= \sum\nolimits_
{\lbrace \lambda\vert m_1-m_2=0 mod3\rbrace}\vert \chi_\lambda\vert ^2
+ \sum\nolimits_{\lbrace \lambda\vert m_1-m_2=2k mod3\rbrace}
\chi_\lambda \bar\chi_{\sigma(\lambda)}\cr
&\> +\sum\nolimits_{\lbrace \lambda\vert m_1-m_2=k mod 3\rbrace}\chi_\lambda
\bar\chi_{\sigma^2 (\lambda)},&(3.27)\cr}$$
where $\sigma (\lambda) = m_2\, \omega_1+(m_1 -m_2)\, \omega_2 $. The
eq. (3.27) agrees with the result of Ref. 17.
\footnote *{There is a minus sign error in Ref. 17. Defining
$\sigma (\lambda )$ the same as in our case, the two terms in the
exponential of eq. (6.4) of Ref. 17 must both have negative signs, in
order for $M_ {\lambda ,\lambda'}$ to commute with the operator T.}

\

\noindent {\bf E- Series}

 We expect to find nondiagonal partition functions at levels $k =5, 9, 21$,
  due to the following conformal embeddings $^{20}$
$$\widehat {su(3)}_{k=3} \subset \widehat {(D_4)}_{k=1}\quad,\quad
\widehat {su(3)}_{k=5} \subset \widehat {su(6)}_{k=1}$$
$$\widehat {su(3)}_{k=9} \subset \widehat {(e_6)}_{k=1}\quad,\quad
\widehat {su(3)}_{k=21} \subset \widehat {(e_7)}_{k=1}.\eqno(3.28)$$
Thus, we begin from these levels.

{\bf 1.}\ At level $k = 3\;(h=6)$, we start with $A_6$ and mod it out by
 {\it allowed} $\caz_N$'s up to $N=18$, but all of them results only in
 a combination of $A_6$ and $D_6$. For example, in the case $N=6$ after
 doing the sum according to eq. (A.5) we obtain
$$\Bigl[\bigl(\sum_{\alpha=1}^6 T^\alpha+\sum_{\alpha=1}^3T^\alpha ST^2
+\sum_{\alpha=1}^2T^\alpha ST^3\bigr)SZ_1+Z_1\Bigr] =3A_6+D_6.\eqno(3.29)$$
Then, we start with $D_6$, mod it out by {\it allowed} moddings but also
nothing more is found. For example, in modding by $\caz_6$ we find
$$\Bigl[\bigl(\sum_{\alpha=1}^6 T^\alpha+\sum_{\alpha=1}^3T^\alpha ST^2
+\sum_{\alpha=1}^2T^\alpha ST^3\bigr)SZ_1+Z_1\Bigr] ={9\over2}D_6.
\eqno(3.30)$$
This is not surprising, since it can easily be seen that $D_6$ given by
$$D_6=\big\vert\chi_{1,1}+\chi_{1,4}+\chi_{4,1}\big\vert^2
+3\big\vert\chi_{2,2}\big\vert^2,\eqno(3.31)$$
exactly corresponding to conformal embedding
$\widehat {su(3)}_{k=3} \subset \widehat {(D_4)}_{k=1}$.

{\bf 2.}\  At level $k=5\; (h=8)$, starting with $D_8$ and modding by
$\caz_8$ according to eq. (A.6),
we see that the sum of terms in the {\it bracket} leads to the case I
of section 2.2 which after subtracting the known partition functions
$A_8^{c.c}$ and $D_8$ each with mutiplicity 2, we obtain
an exceptional partition function, which is called $E_8^{c.c.}$,
$$\Bigl[\bigl(\sum_{\alpha=1}^8 T^\alpha+\sum_{\alpha=1}^2T^\alpha ST^2
+ST^4\bigr)SZ_1+Z_1\Bigr]=2D_8+2A_8^{c.c.}+E_8^{c.c.}\eqno(3.32)$$
\noindent with
$$\eqalign{E_8^{c.c.}&=\big\vert\chi_{1,1}+\chi_{3,3}\big\vert^2+\big\vert
\chi_{1,4}+\chi_{4,1}\big\vert^2\cr
&\;+\bigl((\chi_{3,1}+\chi_{3,4})\overline{(\chi_{1,3}+\chi_{4,3})}+
(\chi_{2,3}+\chi_{6,1})\overline{(\chi_{3,2}+\chi_{1,6})}+c.c.\bigr)
;\cr}\eqno(3.33)$$
\noindent Thus,
$$\eqalignno {E_8&=\big\vert\chi_{1,1}+\chi_{3,3}\big\vert^2+\big\vert
\chi_{1,3}+\chi_{4,3}\big\vert^2+\big\vert\chi_{3,1}+\chi_{3,4}\big\vert^2\cr
&\;+\big\vert\chi_{1,4}+\chi_{4,1}\big\vert^2+
\big\vert\chi_{2,3}+\chi_{6,1}\big\vert^2+\big\vert\chi_{3,2}+\chi_{1,6}
\big\vert^2,&(3.34)\cr}$$
\noindent where $\chi_{(m_1,m_2)}$ denotes the character of HW representation
$\lambda=m_1\omega_1+m_2\omega_2$.
 One can easily see that $E_8$ corresponds to conformal
embedding $\widehat {su(3)}_{k=5}\subset \widehat {su(6)}_{k=1} $.$^{13}$
 It must be mentioned that we also find $E_8$ in modding by
 $\caz_2$ and $\caz_4$, but in these cases we encounter the
 case II of Subsec. 2.2., and we obtain
$$\eqalignno {\Bigl[\sum_{\alpha=1}^2 T^\alpha SZ_1+Z_1\Bigr] &={1\over 2}
\bigl(5D_8-A_8^{c.c.}+E_8\bigr)&(3.35)\cr
 \Bigl[\bigl(\sum_{\alpha=1}^4 T^\alpha+ST^2\bigr)
SZ_1+Z_1\Bigr] & ={1\over 2}\bigl(7D_8+A_8^{c.c.}-E_8^{c.c.}\bigr).&(3.36)
\cr}$$
Modding by $\caz_3$ gives $D_8$ itself, but modding for example by $\caz_5$
or $\caz_7$ are not {\it allowed} because the condition (2.8) is not
satisfied. We have continued modding, up to $\caz_{48}$ but no other
exceptional partition function is found.

{\bf 3.} \  At level $k=9\; (h=12)$, we start with $D_{12}$ and mod it out by
$\caz_9$ and do the sum according to the {\it bracket} of eq. (A.7) with
$N=9$, finally we encounter the case II of Subsec. 2.2., which after
subtraction $D_{12}$  of multiplicity 12, we obtain an exceptional partition
function, denoted by $E_{12}^{(1)}$ with an overal multiplicity $-3$:
$$\Bigl[\bigl(\sum_{\alpha=1}^9 T^\alpha+ST^3+ST^6\bigr)
SZ_1+Z_1\Bigr] =12D_{12}-3E_{12}^{(1)},\eqno(3.37)$$
\noindent where
$$E_{12}^{(1)}=\big\vert\chi_{1,1}+\chi_{1,10}+\chi_{10,1}
+\chi_{2,5}+\chi_{5,2}+\chi_{5,5}\big\vert^2+
2\big\vert\chi_{3,3}+\chi_{3,6}+\chi_{6,3}\big\vert^2.\eqno(3.38)$$
This partition function corresponds to conformal embedding
$\widehat {su(3)}_{k=9} \subset \widehat {(e_6)}_{k=1}$.$^{13}$
In modding $D_{12}$ by $\caz_2$, after doing the sum in the {\it bracket}
of eq. (2.7) with $N=2$, one encounters case II, however in this case the
trace of another modular invariant can easily be seen. Actually subtracting
$D_{12}$, its charge conjugation counterpart $D_{12}^{c.c}$ and
$E_{12}^{(1)}$ with multiplicities $5/2,-1/2$, and $1/2$ respectively, we
obtain another exceptional partition function which we denote by
$E_{12}^{(2)}$
$$\Bigl[\sum_{\alpha=1}^2 T^\alpha SZ_1+Z_1\Bigr]
={1\over2}\bigl(5D_{12}-D_{12}^{c.c.}+E_{12}^{(1)}+E_{12}^{(2)}\bigr),
\eqno(3.39)$$
\noindent with
$$\eqalign {E_{12}^{(2)}=& \big\vert\chi_{1,1}+\chi_{1,10}+\chi_{10,1}
\big\vert^2+\big\vert\chi_{2,5}+\chi_{5,2}+\chi_{5,5}\big\vert^2+
 \big\vert\chi_{3,3}+\chi_{3,6}+\chi_{6,3}\big\vert^2\cr
 &+\big\vert\chi_{1,4}+\chi_{4,7}+\chi_{7,1}\big\vert^2
+\big\vert\chi_{4,1}+\chi_{7,4}+\chi_{1,7}\big\vert^2\cr
&+ 2\big\vert\chi_{4,4}\big\vert^2+\bigl( (\chi_{4,4})\overline{(\chi_{2,2}
+\chi_{2,8}+\chi_{8,2})}+c.c.\bigr),\cr}\eqno(3.40)$$
which does not correspond with a conformal embedding; and was found using
a nontrivial automorphism of the fusion rules of the extended algebra.$^{24}$
We continued modding up to $\caz_{72}$, but no other exceptional theory
appears at this level.

{\bf 4.} \ At level $k = 21\;(h=24)$, starting with $D_{24}$ and modding  by
$\caz_2$ leads to the case III of Subsec. 2.2., which after subtracting
$D_{24}$ of multiplicity 2, yields a modular invariant combination with
some of its coefficients negative integers, which we call $M_{24}$
$$\Bigl[\sum_{\alpha=1}^2 T^\alpha SZ_1+Z_1\Bigr]=2D_{24}+M_{24},\eqno(3.41)
$$
 \noindent with
 $$\eqalign {M_{24}=\biggl[& \Big\vert\chi_{[1,1]}+\chi_{[2,11]}\Big\vert^2
+\Big\vert\chi_{[5,5]}+\chi_{[7,7]}\Big\vert^2 +\Big\vert\chi_{[1,7]}
+\chi_{[8,5]}\Big\vert^2
+\Big\vert\chi_{[7,1]}+\chi_{[5,8]}\Big\vert^2\cr
&+\Big\vert\chi_{[3,3]}+\chi_{[6,9]}\Big\vert^2
+\Big\vert\chi_{[1,4]}-\chi_{[4,7]}\Big\vert^2
+\Big\vert\chi_{[4,1]}-\chi_{[7,4]}\Big\vert^2
+2\Big\vert\chi_{[3,9]}\Big\vert^2\cr
&+2\Big\vert\chi_{[9,3]}\Big\vert^2
 -\Bigl(\big\vert\chi_{[2,2]}+\chi_{[2,8]}+\chi_{[8,2]}-\chi_{[4,10]}
\big\vert^2+3\big\vert\chi_{[4,4]}-\chi_{[8,8]}\big\vert^2 \Bigr)\biggr],
\cr}\eqno(3.42)$$
\noindent where
$$ \chi_{[m_1,m_2]}\equiv\chi_{(m_1,m_2)}
+\chi_{(m_2,h-m_1-m_2)}+\chi_{(h-m_1-m_2,m_1)}.\eqno(3.43)$$
Modding by $\caz_3$ gives $D_{12}$ itself, and modding by $\caz_4$, $\caz_8$,
$\caz_9$, $\caz_{18}$, $\caz_{24}$, $\caz_{36}$ give rise to three extra
modular invariant combinations, which we do not mention here their explicit
form. Finally, modding by $\caz_{72}$ and doing the sum in the {\it bracket}
of eq. (A.16), we encounter the case II of Subsec. 2.2., which after
Subtracting $D_{24}$ and $M_{24}$ and their charge conjugation counterparts
 $D_{24}^{c.c}$, $M_{24}^{c.c}$ with multiplicities 12 and 3 respectively,
 we are left with an exceptional partition function which is denoted by
 $E_{24}$:
$$\eqalignno {\biggl[ & \Bigl(\sum_{\alpha=1}^{72}T^{\alpha}+\sum_{\alpha=1}
^{18}T^{\alpha}ST^2+\sum_{\alpha=1}^8 T^{\alpha}ST^3
+\sum_{\alpha=1}^9 T^{\alpha}ST^4
+\sum_{\alpha=1}^2 T^{\alpha}ST^6 \cr
& \> +\sum_{\alpha=1}^9 T^{\alpha}ST^8
+\sum_{\alpha=1}^8 T^{\alpha}ST^9
+ST^{12}+\sum_{\alpha=1}^8 T^{\alpha}ST^{15}
+\sum_{\alpha=1}^2 T^{\alpha}ST^{18}\cr
& \> +ST^{24}+\sum_{\alpha=1}^2 T^{\alpha}ST^{30}+ST^{36}
+ST^{48}+ST^{60}\Bigr)SZ_1+Z_1\biggr]\cr
&=12D_{24}+12D_{24}^{c.c}+3M_{24}+3M_{24}^{c.c}+9 E_{24},&(3.44)\cr}$$
\noindent with
$$E_{24}=\big\vert\chi_{[1,1]}+\chi_{[5,5]}+\chi_{[2,11]}+\chi_{[7,7]}
\big\vert^2+\big\vert\chi_{[1,7]}+\chi_{[7,1]}+\chi_{[5,8]}+\chi_{[8,5]}
\big\vert^2.\eqno(3.45)$$
This theory corresponds to  conformal embedding
$\widehat {su(3)}_{k=21} \subset \widehat {(e_7)}_{k=1}$.$^{13}$

So with our method we reproduce not only  exceptional partition functions
 corresponding to a certain conformal embedding like $E_8,\, E_{12}^{(1)}$,
 and $E_{24}$, but also the one which can not be obtained by a conformal
 embedding i.e. $E_{12}^{(2)}$. Note that at each level the {\it allowed}
 modding $\caz_N$ has $N$ a divisor of $3h$, where $h=k+3$.

\

\noindent {\bf 3.3. $\widehat {su(4)}$  WZW models}

\noindent In addition to the diagonal series
$A_h = \sum _{\lbrace\lambda \in B_h \rbrace} \vert \chi_\lambda \vert ^2$,
 where
$$B_h = \Bigl\lbrace \lambda\;\big\vert \;  3\leq\sum_{i=1}^3 m_i <
 {h = k + 4}\Bigr\rbrace,\eqno(3.46)$$
and $\lambda =\Sigma_{i=1}^3 m_i \omega_i$,
there exist two $D$ series corresponding to the two subgroups of the centre
of $SU(4)$. Furthermore, up to now three exceptional partition functions have
been found in levels $k=4, 6, 8$. In the following we obtain all of these by
orbifold approach. We define the action of the $\caz_N$ on the characters
of left$-$moving HW representations of $su(4)$ due to eq. (2.9) by
$$p \;\cdot \; \chi_{(m_1,m_2,m_3)}=e^{{2\pi i\over N}
(\varphi_{_\lambda}-20)}\;\chi_{(m_1,m_2,m_3)}\eqno(3.47)$$
where
$$\varphi_{_\lambda} =3m_1\,^2+4m_2\,^2+3m_3\,^2 +4m_1m_2+2m_1m_3+4m_2m_3$$
and $p$ is the generator of $\caz_N$.
Thus, the untwisted part of a partition function, consists of left$-$moving
HW representations $\lambda$ which satisfy: $\varphi_{_\lambda}=20\, mod\,N$.

\

\noindent {\bf D - Series}

 We follow the same recipe of calculation that was described in Subsec. 3.2.,
 but without going into the details, and find the general form of $D_h$
 series at each level.
 Starting with $A_h$, first we mod it out by a subgroup $\caz_2$ of the
 centre and obtain at every level a nondiagonal partition
function which we denote  by $D_h^{(2)}$,
$$D_h^{(2)} \equiv Z(A_h/\caz_2)= \sum\nolimits_{\lbrace
\lambda\vert \Sigma_{i=1}^3 i m_i=0 mod2\rbrace}\vert
\chi_ \lambda\vert ^2 +\sum\nolimits_{\lbrace\lambda\vert \Sigma_{i=1}^3
 i m_i=k mod 2\rbrace } \chi_\lambda \bar\chi_{\mu (\lambda)},\eqno(3.48)$$
\noindent where $\mu (m_1, m_2, m_3) =(m_3, h-\Sigma_{i=1}^3 m_i
, m_1)$.
Then we mod out the $A_h$ series by $\caz_4$ according to eq. (A.4) and find
at every {\it even} level a nondiagonal partition function $D_h^{(4)}$,
which has the form
$$\eqalign {D_h^{(4)}\!\! \equiv Z(A_h/{\caz_4}) &\!\! =\!\! \sum_
{\lbrace\lambda\vert \Sigma_{i=1}^3 i m_i =2\, mod\, 4\rbrace}\quad\vert
\chi_{\lambda}\vert ^2\quad +\! \sum_{\lbrace\lambda\vert \Sigma_{i=1}^3
 i m_i =2+{k\over 2}\, mod\, 4\rbrace}\! \chi_{\lambda} \bar
 \chi_{\sigma(\lambda)}\cr
& \, +\!\! \sum_{\lbrace\lambda\vert \Sigma_{i=1}^3 i m_i =2-k\, mod\, 4
\rbrace} \chi_{\lambda} \bar\chi_{\sigma^2 (\lambda)}
+\!\!\sum_{\lbrace\lambda\vert \Sigma_{i=1}^3 i m_i =2-{k\over 2} mod 4
\rbrace}\chi_{\lambda} \bar\chi_{\sigma^3 (\lambda)},\cr}\eqno(3.49)$$
 where $\sigma (m_1, m_2, m_3)=(m_2, m_3, h -\Sigma_{i=1}^3 m_i)$.
 These results agree with the ones obtained in Ref. 17 modulo the comment
 in the footnote of page 13.

\

\noindent {\bf E - Series}

 It is expected that there are nondiagonal partition functions at levels
 $k =2,4,6,8$, due to the following conformal embeddings:$^{20}$
$$\widehat {su(4)}_{k=2} \subset \widehat {su(6)}_{k=1}\quad,
\quad\widehat {su(4)}_{k=4} \subset \widehat {(B_7)}_{k=1}$$
$$\widehat {su(4)}_{k=6} \subset \widehat {su(10)}_{k=1}\quad,
\quad\widehat {(D_3)}_{k=8} \subset \widehat {(D_{10})}_{k=1}.\eqno(3.50)$$

{\bf 1.}\ At level $k = 2 \; (h = 6)$, we have only $D_6^{(2)}$,
 and $D_6^{(4)}= A_6^{c.c.}$. First, we start with $A_6$ and mod it out by
 all {\it allowed} $\caz_N$'s up to $N=24$. Nothing other
than $A_6$ and $D_6^{(2)}$ and their charge cojugation counterparts
is found. For example, in the cases $N=3,6,8$ we obtain
$$\eqalignno {\Bigl[\sum_{\alpha=1}^3 T^\alpha SZ_1+Z_1\Bigr] &
=4A_6-2D_6^{(2)}\;\;\;&(3.51)\cr
\Bigl[\bigl(\sum_{\alpha=1}^6 T^\alpha+\sum_{\alpha=1}^3T^\alpha ST^2+
\sum_{\alpha=1}^2T^\alpha ST^3\bigr)SZ_1+Z_1\Bigr] & =4A_6+4D_6^{(2)}
&(3.52)\cr\Bigl[\bigl(\sum_{\alpha=1}^8 T^\alpha
+\sum_{\alpha=1}^2T^\alpha ST^2
+ST^4\bigr)SZ_1+Z_1\Bigr] & =2A_6+2A_6^{c.c.}+D_6^{(2)},&(3.53)\cr}$$
respectively. Then we start with $D_6^{(2)}$ and do the {\it allowed}
moddings. Again, nothing more is found. For example, in modding by
$\caz_3$ we find
$$\Bigl[\sum_{\alpha=1}^3 T^\alpha SZ_1+Z_1\Bigr]=2D_6^{(2)}.\eqno(3.54)$$
This is not surprising, since it can easily be seen that $D_6^{(2)}$
$$D_6^{(2)}=\big\vert\chi_{(1,1,1)}+\chi_{(1,3,1)}\big\vert^2
+\big\vert\chi_{(1,1,3)}+\chi_{(3,1,1)}\big\vert^2
+2\big\vert\chi_{(1,2,1)}\big\vert^2+2\big\vert\chi_{(2,1,2)}\big\vert^2
,\eqno(3.55)$$
exactly corresponding to conformal embedding
$\widehat {su(4)}_{k=2} \subset \widehat {su(6)}_{k=1}.$

{\bf 2.}\ At level $k = 4 \; (h=8)$, there exist two $D_8^{(2)}$ and
$D_8^{(4)}$ partition functions.
We choose to start with $D_8^{(2)}$. Modding by $\caz_2$
gives $D_8^{(2)}$ itself, and modding by $\caz_4$ and $\caz_8$ give
a combination of $D_8^{(2)}$ and $D_8^{(4)}$:
$$\eqalignno {\Bigl[\bigl(\sum_{\alpha=1}^4 T^\alpha+ST^2\bigr)
SZ_1+Z_1\Big]& = 2D_8^{(2)}+2D_8^{(4)}&(3.56)\cr
\Bigl[\bigl(\sum_{\alpha=1}^8 T^\alpha+\sum_{\alpha=1}^2T^\alpha ST^2
+ST^4\bigr)SZ_1+Z_1\Bigr] & =4D_8^{(2)}+4D_8^{(4)}.&(3.57)\cr}$$
The next modding which satisfies the condition (A.3) is $\caz_{16}$.
After doing the sum in the {\it bracket} of eq. (A.10) we encounter
the case I of Subsec. 2.2., which after subtracting the
$D_8^{(2)}$ and $D_8^{(4)}$ each with multiplicity $4$, an exceptional
partition function is found which we denote by $E_8$:
$$\Bigl[\bigl(\sum_{\alpha=1}^{16} T^\alpha+\sum_{\alpha=1}^4T^\alpha ST^2
+ST^4+ST^8+ST^{12}\bigr)SZ_1+Z_1\Bigr]=4D_8^{(2)}+4D_8^{(4)}+4E_8,
\eqno(3.58)$$
\noindent where
$$\eqalignno {E_8 & =\big\vert\chi_{(1,1,1)}+\chi_{(1,5,1)}
+\chi_{(1,2,3)}+\chi_{(3,2,1)}\big\vert^2\cr
& \> +\big\vert\chi_{(1,1,5)}+\chi_{(5,1,1)}+
\chi_{(2,1,2)}+\chi_{(2,3,2)}\big\vert^2
+4\big\vert\chi_{(2,2,2)}\big\vert^2.&(3.59)\cr}$$
It can easily be shown that this partition function corresponds to
conformal embedding $\widehat {su(4)}_{k=4} \subset \widehat {(B_7)}_{k=1}$.
The above exceptional partition function was obtained in the context of
fixed-point resolution of Ref. 19.
 We then repeat the above moddings starting with $D_8^{(4)}$ and
 find exactly the same results as with $D_8^{(2)}$.

{\bf 3.}\ At level $k = 6 \; (h=10)$, there exist $D_{10}^{(2)}$ and
$D_{10}^{(4)}$. Starting with $D_{10}^{(2)}$ the following results are
obtained. Modding by $\caz_2$ and $\caz_4$ gives $D_{10}^{(2)}$ itself, but
modding by $\caz_8$ and doing the sum in the {\it bracket} of eq. (A.6) we
encounter the case III of Subsec. 2.2., which after subtraction
$D_{10}^{(2)}$ of multiplicity 6 leads to a modular invariant combination,
which we call $M_{10}$
$$\Bigl[\bigl(\sum_{\alpha=1}^8 T^\alpha+\sum_{\alpha=1}^2T^\alpha ST^2
+ST^4\bigr)SZ_1+Z_1\Bigr]=6D_{10}-M_{10}\eqno(3.60)$$
\noindent where
$$\eqalignno {M_{10} & =\,\big\vert(\chi_{(1,1,1)}+\chi_{(1,7,1)})
-(\chi_{(2,1,6)}+\chi_{(6,1,2)})-(\chi_{(3,1,3)}+\chi_{(3,3,3)})\big\vert^2
\cr& \> +\big\vert(\chi_{(1,1,7)}+\chi_{(7,1,1)})
-(\chi_{(1,2,1)}+\chi_{(1,6,1)})-(\chi_{(1,3,3)}+\chi_{(3,3,1)})\big\vert^2
\cr&+ 3 \big\vert\chi_{(2,2,4)}+\chi_{(4,2,2)}\big\vert^2
+ 3 \big\vert\chi_{(2,2,4)}+\chi_{(4,2,2)}\big\vert^2.&(3.61)\cr}$$
In modding by $\caz_5$ after doing the sum in the {\it bracket} of eq. (2.7)
with $N=5$, we encounter the case II of Subsec. 2.2., which by subtracting
$D_{10}^{(2)}$, and its charge conjugation
$D_{10}^{(2) \; c.c.}$, and $M_{10}$ by multiplicities $-3/5,4/5$, and $8/5$
respectively, an exceptional partition function is found which we call
$E_{10}$
$$\Bigl[\sum_{\alpha=1}^5 T^\alpha SZ_1+Z_1\Bigr]
={1\over5}\bigl(4D_{10}^{(2)\,c.c.}-3D_{10}^{(2)}+8M_{10}+2E_{10}\bigr),
\eqno(3.62)$$
\noindent where
$$\eqalign {E_{10} & = \,\big\vert\chi_{(1,1,1)}+\chi_{(1,7,1)}
+\chi_{(3,1,3)}+\chi_{(3,3,3)}\big\vert^2\cr
& \; +\big\vert\chi_{(1,1,7)}+\chi_{(7,1,1)}+\chi_{(1,3,3)}+\chi_{(3,3,1)}
\big\vert^2\cr
& \; +\big\vert\chi_{(1,1,3)}+\chi_{(3,5,1)})+\chi_{(3,2,3)}\big\vert^2
+\big\vert\chi_{(3,1,1)}+\chi_{(1,5,3)})+\chi_{(3,2,3)}\big\vert^2\cr
& \> +\big\vert\chi_{(1,1,5)}+\chi_{(5,3,1)})+\chi_{(2,3,2)}\big\vert^2
+\big\vert\chi_{(5,1,1)}+\chi_{(1,3,5)})+\chi_{(2,3,2)}\big\vert^2\cr
& \> +\big\vert\chi_{(1,2,3)}+\chi_{(3,4,1)})+\chi_{(4,1,4)}\big\vert^2
+\big\vert\chi_{(1,4,3)}+\chi_{(3,2,1)})+\chi_{(4,1,4)}\big\vert^2\cr
& \; +\big\vert\chi_{(2,1,4)}+\chi_{(4,3,2)})+\chi_{(1,4,1)}\big\vert^2
+\big\vert\chi_{(4,1,2)}+\chi_{(2,3,4)})+\chi_{(1,4,1)}\big\vert^2.
\cr}\eqno(3.63)$$
It can easily be shown that $E_{10}$ actually corresponds to
conformal embedding
$\widehat {su(4)}_{k=6}\subset \widehat {su(10)}_{k=1}$.
 This exceptional partition function was found in the context of
simple current method in Ref. 18. We have carried out all {\it allowed}
moddings up to $\caz_{40}$, and except in the case of $\caz_8$ which
gives rise to another modular invariant combibation, we find nothing
new other than some linear combination
of $D_{10}^{(2)}$, $M_{10}$, and $E_{10}$ and their charge cojugations. For
example, modding by $\caz_{40}$ according to eq. (A.15)  yields
$$\eqalignno {\Bigl[ & \bigl(\sum_{\alpha=1}^{40}T^{\alpha}+\sum_{\alpha=1}
^{20}T^{\alpha}ST^2+\sum_{\alpha=1}^5 T^{\alpha}ST^4
+\sum_{\alpha=1}^8 T^{\alpha}ST^5\cr
& \> +\sum_{\alpha=1}^5 T^{\alpha}ST^8
 +\sum_{\alpha=1}^2 T^{\alpha}ST^{10}+ST^{20}\bigr)SZ_1+Z_1\Bigr]\cr
&\> ={6\over5}\Bigl(2D_{10}^{(2)}+4D_{10}^{(2)\,c.c.}+3M_{10}+2E_{10}
\Bigr)&(3.64)\cr}$$

{\bf 4.}\ At level $k =8 \; (h = 12)$, there exist $D_{12}^{(2)}$ and
$D_{12}^{(4)}$.
We choose $D_{12}^{(4)}$ and obtain the following results. Modding by
$\caz_2$ and $\caz_4$ gives $D_{12}^{(4)}$ itself, but in modding by $\caz_8$
we ecounter the case II of Subsec. 2.2., which after subtraction
$D_{12}^{(4)}$ of multiplicity $12$, an exceptional partition function
is found, which we call $E_{12}^{(1)}$
$$\Bigr[\bigl(\sum_{\alpha=1}^8 T^\alpha+\sum_{\alpha=1}^2T^\alpha ST^2
+ST^4\bigr)SZ_1+Z_1\Bigr]=12D_{12}^{(4)}-2E_{12}^{(1)},\eqno(3.65)$$
\noindent where
\footnote *{We are not aware of the explicit form of this partition
function in the literature.}

$$\eqalign {E_{12}^{(1)}\!=&\big\vert\chi_{(1,1,1)}\!+\!\chi_{(1,1,9)}
\!+\!\chi_{(1,9,1)}\!+\!\chi_{(9,1,1)}\!+\!\chi_{(2,3,2)}\!+\!\chi_{(2,5,2)}
\!+\!\chi_{(3,2,5)}\!+\!\chi_{(5,2,3)}\big\vert^2\cr
& \, +\!\big\vert\chi_{(1,3,1)}\!+\!\chi_{(1,7,1)}
\!+\!\chi_{(3,1,7)}\!+\!\chi_{(7,1,3)}\!+\!\chi_{(1,4,3)}\!+\!\chi_{(3,4,1)}
\!+\!\chi_{(4,1,4)}\!+\!\chi_{(4,3,4)}\big\vert^2\cr
& \, +2\big\vert\chi_{(2,2,4)}\!+\!\chi_{(2,4,4)}\!+\!\chi_{(4,2,2)}
+\chi_{(4,4,2)}\big\vert^2.\cr}\eqno(3.66)$$
It can easily be seen that $E_{12}^{(1)}$ just corresponds to conformal
embedding
$\widehat {(D_3)}_{k=8}\subset \widehat {(D_{10})}_{k=1}$ with the following
branching rules:
$$\eqalignno {ch_1& = \chi_{(1,1,1)}\!+\!\chi_{(1,1,9)}\!+\!\chi_{(1,9,1)}
\!+\!\chi_{(9,1,1)}\!+\!\chi_{(2,3,2)}\!+\!\chi_{(2,5,2)}\!+\!\chi_{(3,2,5)}
\!+\!\chi_{(5,2,3)}\cr
ch_2& =\chi_{(1,3,1)}\!+\!\chi_{(1,7,1)}\!+\!\chi_{(3,1,7)}\!
+\!\chi_{(7,1,3)}\!+\!\chi_{(1,4,3)}\!+\!\chi_{(3,4,1)}\!
+\!\chi_{(4,1,4)}\!+\!\chi_{(4,3,4)}\cr
ch_3& =ch_4=\chi_{(2,2,4)}\!+\!\chi_{(2,4,4)}\!+\!\chi_{(4,2,2)}\!
+\!\chi_{(4,4,2)},&(3.67)\cr}$$

 where $chi_i$'s  are the  characters  of the integrable
representations of $(\widehat {(D_{10})}_{k=1}$
and $\chi_{(m_1,m_2,m_3)}$'s are those of $\widehat {SU(4)}_{k=8}$.

Then, modding by $\caz_3$ and doing the sum in eq. (2.7) with
 $N=3$ leads to the case I of Subsec. 2.2., which after subtraction
 $D_{12}$, $D_{12}^{c.c.}$, and $E_{12}^{(1)}$ each of multiplicity $1/3$
 another exceptional partition function is found, which we call
 $E_{12}^{(2)}$
$$\Bigl[\bigl(\sum_{\alpha=1}^4 T^\alpha+ST^2\bigr)
SZ_1+Z_1\Bigr]={1\over 3}\Bigl( D_{12}^{(4)}+D_{12}^{(4)\> c.c.}
+E_{12}+4E_{12}^{(2)}\Bigr)\eqno(3.68)$$
\noindent with
$$\eqalign {E_{12}^{(2)}&\!\! =\!\! \big\vert\chi_{(1,1,1)}\!+\!
\chi_{(1,1,9)}\!+\!\chi_{(1,9,1)}
\!+\!\chi_{(9,1,1)}\big\vert^2\!+\!\big\vert\chi_{(2,3,2)}\!+\!\chi_{(2,5,2)}
\!+\!\chi_{(3,2,5)}\!+\!\chi_{(5,2,3)}\big\vert^2\cr
& +\!\!\big\vert\chi_{(1,3,1)}\!+\!\chi_{(1,7,1)}\!+\!\chi_{(3,1,7)}\!
+\!\chi_{(7,1,3)}\big\vert^2\!+\!\big\vert\chi_{(1,4,3)}\!+\!\chi_{(3,4,1)}
\!+\!\chi_{(4,1,4)}\!+\!\chi_{(4,3,4)}\big\vert^2\cr
& +\!\!\big\vert\chi_{(1,1,5)}\!+\!\chi_{(1,5,5)}\!+\!\chi_{(5,1,1)}
\!+\!\chi_{(5,5,1)}\big\vert^2\!+\!\big\vert\chi_{(1,3,5)}\!+\!\chi_{(3,1,3)}
\!+\!\chi_{(3,5,3)}\!+\!\chi_{(5,3,1)}\big\vert^2\cr
& +\!\!\big\vert\chi_{(1,5,1)}\!+\!\chi_{(5,1,5)}\big\vert^2\!
+\big\vert\chi_{(2,2,4)}\!+\!\chi_{(2,4,4)}\!+\!\chi_{(4,2,2)}\!
+\!\chi_{(4,4,2)}\big\vert^2\!+\!2\big\vert\chi_{(3,3,3)}\big\vert^2\cr
& +\!\!\biggl(\left(\chi_{(1,2,3)}\!+\!\chi_{(1,6,3)}\!
+\!\chi_{(2,1,6)}\!+\!\chi_{(2,3,6)}\!+\!\chi_{(3,2,1)}\!+\!\chi_{(3,6,1)}
\!+\!\chi_{(6,1,2)}+\!\chi_{(6,3,2)}\right)\cr
&\quad\cdot\overline{\chi_{(3,3,3)}}\!+\!(\chi_{(1,5,1)}\!
+\!\chi_{(5,1,5)})\left(\overline{\chi_{(1,2,7)}\!+\!\chi_{(2,1,2)}\!
+\!\chi_{(2,7,2)}\!+\!\chi_{(7,2,1)}}\right)\!+\!c.c.\!\biggr).\cr}
\eqno(3.69)$$
This exceptional partition function which doesn't correspond to a conformal
emedding, was recently found by a computational method which essentially
looks for the eigenvectors of matrix S (the generator of the modular group)
with eigenvalues equal to one,$^{25,28}$ and can be shown to be a
consequence of an automorphism of the fusion rules of the extended algebra.$^
{28}$ We continue the modding by {\it allowed} groups up to
$\caz_{96}$, but no new partition function is found.
 For example in modding by $\caz_{24}$ according to eq. (A.12), we get
$$\eqalign {\Bigl[ & \bigl(\sum_{\alpha=1}^{24}T^{\alpha}+\sum_{\alpha=1}
^{6}T^{\alpha}ST^2+\sum_{\alpha=1}^8 T^{\alpha}ST^3
+\sum_{\alpha=1}^3 T^{\alpha}ST^4
+\sum_{\alpha=1}^2 T^{\alpha}ST^6 \cr
& \> +\sum_{\alpha=1}^3 T^{\alpha}ST^8+ST^{12}\bigr)SZ_1+Z_1\Bigr]
=4\bigl(D_{12}+D_{12}^{c.c}+E_{12}^{(1)}+4E_{12}^{(2)}\bigr).\cr}
\eqno(3.70)$$
So again, as in the case of $\widehat {su(3)}$ models, with the orbifold
method not only exceptional partition functions which correspond to
conformal embeddings like $E_8,\, E_{10}$, and $E_{12}^{(1)}$ are found,
but also the a partition function which does not corresponds to a
conformal embedding i.e. $E_{12}^{(2)}$, is generated.

In this subsection we have been able to obtain explicitly all the
exceptional partition functions which correspond to a conformal embedding and
moreover
the one ($E_{12}^{(2)}$) which follows from an automorphism of the
fusion rules of the extended algebra.
 Note that at each level the {\it allowed}
 modding $\caz_N$ has $N$ a divisor of $4h$, where $h=k+4$.

\

\noindent {\bf 3.4. $\widehat {su(5)}$ WZW models}

\noindent For $\widehat {su(5)}$  models besides the usual diagonal series
$A_h = \sum _{\lbrace\lambda \in B_h \rbrace} \vert \chi_\lambda \vert ^2$,
 at each level, where
$$B_h = \Bigl\lbrace \lambda\;\big\vert \; 4\leq\sum_{i=1}^4 m_i <
{h = k + 5}\Bigr\rbrace,\eqno(3.71)$$
and $\lambda=\Sigma_{i=1}^4m_i\omega_i$, and one nondiagonal $D$ series at
each level, up to now some exceptional partition functions have been found
which we are going to obtain by our method. We define the action of the
$\caz_N$ on the characters of HW representations of $su(5)$ due to eq. (2.9)
by
$$\eqalign {p \;\cdot\; \chi_{(m_1,m_2,m_3,m_4)}= e^{{2\pi i\over N}
(\varphi_{_\lambda}-50)}\;\chi_{(m_1,m_2,m_3,m_4)},\cr}\eqno(3.72)$$
where
$$\eqalign{\varphi_{_\lambda} =&4m_1^2\!+\!6m_2^2\!+\!6m_3^2\!+\!4m_4^2\!
+\!6m_1m_2\!+\!4m_1m_3\!\cr
&+\!2m_1m_4\!+\!8m_2m_3\!+\!4m_2m_4\!+\!6m_3m_4,\cr}$$
\noindent and $p$ is the generator of $\caz_N$.
 Thus, the untwisted part of a partition function, consists of left$-$moving
 HW representations $\lambda$ which satisfy:
 $\varphi_{_\lambda} = 50\> mod \>N$.

\

\noindent {\bf D - Series}

We start with  $A_h$ series, following the same recipe mentioned in
Subsec. 3.2., mod it out by $\caz_5$ using the eq. (2.7) with $N=5$.
Doing the sum in the {\it bracket}, we obtain the general form of
$D_h$ series:
$$\eqalignno {D_h \equiv Z(A_h/{\caz_5}) & = \sum\nolimits_
{\lbrace\lambda\vert \Sigma_{i=1}^4 i m_i =0 mod 5\rbrace}\quad\vert\chi_
{\lambda}\vert ^2\!+\! \sum\nolimits_{\lbrace\lambda\vert
\Sigma_{i=1}^4 i m_i=3k mod 5\rbrace} \chi_{\lambda} \bar\chi_
{\sigma(\lambda)}\cr
&\> +\!\! \sum\nolimits_{\lbrace\lambda\vert \Sigma_{i=1}^4 i m_i =
k mod 5\rbrace}\!\! \chi_{\lambda} \bar\chi_{\sigma^2 (\lambda)}
\!\!+\!\!\sum\nolimits_{\lbrace\lambda\vert \Sigma_{i=1}^4 i m_i =4k mod
5\rbrace}
\!\!\chi_{\lambda} \bar\chi_{\sigma^3 (\lambda)}\cr
& \> +\!\sum\nolimits_{\lbrace\lambda\vert \Sigma_{i=1}^4 i m_i =2k mod
5\rbrace}
\!\! \chi_{\lambda} \bar\chi_{\sigma^4 (\lambda)}&(3.73)\cr}$$

\noindent where $\sigma (m_1, m_2, m_3, m_4)$ = $(m_2, m_3, m_4,
h -\Sigma_{i=1}^4 m_i)$. These results agree with the ones obtained in
Ref. 17, modulo the comment mentioned in the footnote of page 13.

\

\noindent {\bf E - Series}

One expects to find, possibly, the exceptional series at levels
$k = 3, 5, 7 $
according to the following conformal embeddings:$^{20}$
$$\widehat {su(5)}_{k=3} \subset \widehat {su(10)}_{k=1},\quad
\widehat {su(5)}_{k=5} \subset \widehat {(D_{12}))}_{k=1},\quad \widehat
 {su(5)}_{k=7} \subset \widehat {su(15)}_{k=1}.\eqno(3.74)$$
We will limit ourselves only to the first two cases in this work.

{\bf 1.}\ At level $k = 3\;(h=8)$, we start with $A_8$, doing
the {\it allowed} moddings and obtain the following results.
Modding by $\caz_2$ gives $A_8$ itself, but modding by $\caz_4$
we encounter the case II of section 2.2., which after subtracting $A_8$ and
$D_8^{c.c}$ with multiplicities $5$ and $-1$ respectively, an
exceptional partition function is found which we call $E_8$
$$\Bigl[\bigl(\sum_{\alpha=1}^4 T^\alpha+ST^2\bigr)
SZ_1+Z_1\Bigr]=5A_8-D_8^{c.c.}+E_8\eqno(3.75)$$
\noindent with
$$\eqalignno {E_8 & =
 \big\vert\chi_{(1,1,1,1)}+\chi_{(1,2,2,1)}\big\vert^2
+\big\vert\chi_{(1,1,1,4)}+\chi_{(2,2,1,2)}\big\vert^2\cr
&\quad \>  \big\vert\chi_{(1,1,2,1)}+\chi_{(1,3,1,2)}\big\vert^2
+\big\vert\chi_{(1,1,3,1)}+\chi_{(3,1,1,2)}\big\vert^2\cr
& \quad \>  \big\vert\chi_{(1,1,4,1)}+\chi_{(2,1,2,1)}\big\vert^2
+\big\vert\chi_{(1,2,1,3)}+\chi_{(3,1,2,1)}\big\vert^2\cr
& \quad \>  \big\vert\chi_{(1,2,1,2)}+\chi_{(1,4,1,1)}\big\vert^2
+\big\vert\chi_{(1,3,1,1)}+\chi_{(2,1,1,3)}\big\vert^2\cr
& \quad \>  \big\vert\chi_{(1,2,1,1)}+\chi_{(2,1,3,1)}\big\vert^2
+\big\vert\chi_{(2,1,2,2)}+\chi_{(4,1,1,1)}\big\vert^2.&(3.76)\cr}$$

We have checked that $E_8$ exactly corresponds to conformal
embedding $\widehat {su(5)}_{k=3} \subset \widehat {su(10)}_{k=1}$.
 We carried out all the {\it allowed} moddings up to $\caz_{40}$ and no other
 exceptional partition function was found.
 Then, we start with $D_8$ and mod it out by {\it allowed} moddings up to
 $\caz_{40}$, but again no more exceptional partition function is found.
 For example, modding by $\caz_{16}$ after doing the sum in the {\it bracket}
of eq. (A.10), we encounter the case I of Subsec. 2.2., which after
subtraction $D_8$ and $A_8^{c.c}$ each of multiplicity $4$, results in
 $E_8^{c.c.}$:
$$\Bigl[\bigl(\sum_{\alpha=1}^{16} T^\alpha+\sum_{\alpha=1}^4T^\alpha ST^2
+ST^4+ST^8+ST^{12}\bigr)SZ_1+Z_1\Bigr]=4D_8+4A_8^{c.c.}+2E_8^{c.c.}.
\eqno(3.77)$$

{\bf 2.}\  At level $k = 5\; (h = 10)$, we start with $D_{10}$ which has the
form
$$D_{10} =\big\vert\chi_1\big\vert^2+\big\vert\chi_2\big\vert^2+
\big\vert\chi_3\big\vert^2+\big\vert\chi_4\big\vert^2+
\big\vert\chi_5\big\vert^2+5\big\vert\chi_6\big\vert^2,\eqno(3.78)$$
\noindent here following the Ref. 19, we have used the following
abbreviations
$$\eqalignno {\chi_1& =\chi_{(1,1,1,1)}+\chi_{(1,1,1,6)}+\chi_{(1,1,6,1)}
+\chi_{(1,1,6,1)}+\chi_{(1,6,1,1)}\cr
\chi_2& =\chi_{(1,2,1,3)}+\chi_{(2,1,3,3)}+\chi_{(1,3,3,1)}
+\chi_{(3,3,1,2)}+\chi_{(3,1,2,1)}\cr
 \chi_3& =\chi_{(1,1,2,4)}+\chi_{(1,2,4,2)}+\chi_{(2,4,2,1)}
+\chi_{(4,2,1,1)}+\chi_{(2,1,1,2)}\cr
\chi_4& =\chi_{(1,1,3,2)}+\chi_{(1,3,2,3)}+\chi_{(3,2,3,1)}
+\chi_{(2,3,1,1)}+\chi_{(3,1,1,3)}\cr
\chi_5& =\chi_{(1,2,2,1)}+\chi_{(2,2,1,4)}+\chi_{(2,1,4,1)}
+\chi_{(1,4,1,2)}+\chi_{(4,1,2,2)}\cr
\chi_6& =\chi_{(2,2,2,2)},&(3.79)\cr}$$
 and do the {\it allowed} moddings. Then the following results are
obtained. Modding by $\caz_2$ gives $D_{10}$ itself, but modding by
$\caz_{25}$ according to eq. (A.13), after doing the sum in the
{\it bracket}, we encounter the case I which after subtraction
$D_{10}$ of multiplicity $10$, results in a modular invariant partition
function , which we denote by $E_{10}^{(1)}$:
$$\Bigl[\bigl(\sum_{\alpha=1}^{25} T^\alpha+ST^5+ST^{10}+ST^{15}
+ST^{20}\bigr)SZ_1+Z_1\Bigr]=10D_{10}+10E_{10}^{(1)},\eqno(3.80)$$
\noindent where
$$E_{10}^{(1)} = \big\vert\chi_1 +\chi_2 \big\vert^2
+2 \big\vert\chi_3 \big\vert^2+10 \big\vert\chi_6 \big\vert^2.\eqno(3.81)$$

Next, we mod out $D_{10}$ by $\caz_4$ and encounter case II of Subsec. 2.2.,
which after subtraction $D_{10}$ and $E_{10}^{(1)}$ of mutiplicities $7/2$
and $-3/2$ respetively, leads to another modular invariant partition function
 denoted by $E_{10}^{(2)}$:
$$\Bigl[\bigl(\sum_{\alpha=1}^4 T^\alpha+ST^2\bigr)
SZ_1+Z_1\Bigr]={1\over 2}\bigl(7 D_{10}-3E_{10}^{(1)}+5E_{10}^{(2)}\bigr)
\eqno(3.82)$$
\noindent with
$$E_{10}^{(2)} = \big\vert\chi_1\big\vert^2+\big\vert\chi_2 \big\vert^2
+\big\vert\chi_4 \big\vert^2+\big\vert\chi_5 \big\vert^2
+4\big\vert\chi_6 \big\vert^2+\bigl(\chi_3 \overline{\chi_6}+c.c.\bigr).
\eqno(3.83)$$

What is interesting here is that starting with $E_{10}^{(1)}$ and modding
by $\caz_4$ gives rise to the case II, which after subtracting
$E_{10}^{(1)}$ from the sum in the {\it bracket} of eq. (A.4) with
multiplicity $-2$, we find another modular invariant partition function,
which we denote by $E_{10}^{(3)}$:
$$\Bigl[\bigl(\sum_{\alpha=1}^4 T^\alpha+ST^2\bigr)
SZ_1+Z_1\Bigr]=-2E_{10}^{(1)}+5E_{10}^{(3)},\eqno(3.84)$$
\noindent where
$$E_{10}^{(3)} = \big\vert\chi_1+\chi_2 \big\vert^2
+\big\vert\chi_3 +\chi_6 \big\vert^2+2\big\vert 2 \chi_6 \big\vert^2.
\eqno(3.85)$$
It must be mentioned that modding by $\caz_2$ gives $E_{10}^{(1)}$ itself.

Then we start with $E_{10}^{(2)}$ and do the same moddings. We find that
moddings by $\caz_2$ gives $E_{10}^{(2)}$ itself; but modding by
$\caz_4$ gives rise to the case II which after subtraction of
$E_{10}^{(1)}$, $E_{10}^{(2)}$, and $E_{10}^{(3)}$ of multiplicities
$-1/2,6$,and $1/2$, leads to yet another modular invariant partition
function,which we call $E_{10}^{(4)}$
$$\Bigl[\bigl(\sum_{\alpha=1}^4 T^\alpha+ST^2\bigr)
SZ_1+Z_1\Bigr]={1\over2}\bigl(E_{10}^{(3)}-E_{10}^{(1)}+12E_{10}^{(2)}
-3E_{10}^{(4)}\bigr),\eqno(3.86)$$
\noindent where
$$E_{10}^{(4)} = \big\vert\chi_1+\chi_2 \big\vert^2+8\big\vert\chi_6
\big\vert^2+2\bigl(\chi_3 \overline{\chi_6}+c.c.\bigr).\eqno(3.87)$$
One can easily see that among these  exceptional partition functions,
only $E_{10}^{(3)}$ exactly corresponds to a conformal embedding
$\widehat {su(5)}_{k=5} \subset \widehat {(D_{12}))}_{k=1}$, and
The others were obtained only in the context of the automorphism of fusion
rules techniques.$^{19}$ So in this subsection we have generated by orbifold
method not only the partition functions which can be obtained by
conformal embedding like $E_8$ and $E_{10}^{(3)}$, but also the other
ones which do not correspond to a conformal embedding like
$E_{10}^{(1)}$ , $E_{10}^{(2)}$, and $E_{10}^{(4)}$.

\

\

\noindent {\bf 3.5. $\widehat {su(6)}$ WZW models}

\noindent In addition to the diagonal series $A_h = \sum _{\lbrace\lambda \in
B_h \rbrace} \vert \chi_\lambda \vert ^2$ where
$$B_h = \bigg\lbrace \lambda\, \bigm\vert \; 5\leq \sum_{i=1}^5 m_i
< {h = k + 6}\bigg\rbrace,\eqno(3.88)$$
where $\lambda =\Sigma_{i=1}^5 m_i \omega_i$
there exist three $D_h$ series corresponding to the three subgroups of the
centre of $SU(6)$. So far no exceptional series has been found explicitly
in the literature, however by applying our method we are able to find,
at the first step, a new exceptional partition functions, which could in
principle be interpreted as a conformal embedding.
We define the action of the $\caz_N$ on the characters of HW
representations of $su(6)$ due to eq. (2.9) by
$$p \;\cdot\; \chi_{(m_1,m_2,m_3,m_4,m_5)}= e^{({2\pi i\over N}
\varphi_{_\lambda}-105)}\;\chi_{(m_1,m_2,m_3,m_4,m_5)}\eqno(3.89)$$
where
$$\eqalign {\varphi_{_\lambda}=& 5m_1^2\!+\!8m_2^2\!+\!9m_3^2\!+\!8m_4^2\!
+\!5m_5^2\!+\!8m_1m_2\!+\!6m_1m_3\!+\!4m_1m_4\!+\!2m_1m_5\!\cr
&\, +\!12m_2m_3+\!8m_2m_4\!+\!4m_2m_5+\!12m_3m_4\!+6m_3m_5\!+\!8m_4m_5,\cr}$$
and $p$ is the generator of $\caz_N$.
 Thus the untwisted part of a partition function, consists of left$-$moving
 HW representations $\lambda$ which satisfy:
 $\varphi_{_\lambda}=105 \> mod \> N$.

\

\noindent {\bf D - Series}

Just as in previous cases, we start with $A_h$ and mod it out by
subgroup $\caz_2$ and obtain a nondiagonal series which we denote by
$D_h^{(2)}$:
$$D_h^{(2)} \equiv Z(A_h/\caz_2)= \sum\nolimits_{\lbrace
\lambda\vert \Sigma_{i=1}^5 i m_i=1 mod2\rbrace}\vert
\chi_ \lambda\vert ^2 +\sum\nolimits_{\lbrace\lambda\vert \Sigma_{i=1}^5
i m_i=1+{k\over2} mod 2\rbrace } \chi_\lambda \bar\chi_{\mu (\lambda)}
\eqno(3.90)$$
\noindent for even levels, where
$\mu (m_1, m_2, m_3,m_4,m_5) =(m_4,m_5, h-\Sigma_{i=1}^5 m_i
, m_1,m_2)$.
Then, modding by subgroup $\caz_3$ yields at any level a nondiagonal
partition function, which we call $D_h^{(3)}$
$$\eqalign {D_h^{(3)} \equiv Z(A_h/\caz_3)&= \sum\nolimits_
{\lbrace \lambda\vert \Sigma_{i=1}^5 i m_i=0 mod3\rbrace}\vert \chi_
\lambda\vert ^2 + \sum\nolimits_{\lbrace \lambda\vert \Sigma_{i=1}^5 i m_i
=k mod3\rbrace}\chi_\lambda \bar\chi_{\nu(\lambda)}\cr
&\> +\sum\nolimits_{\lbrace \lambda\vert \Sigma_{i=1}^5 i m_i=2k mod3
\rbrace} \> \chi_\lambda \bar\chi_{\nu^2 (\lambda)}\cr}\eqno(3.91)$$
\noindent where
$\nu (m_1, m_2, m_3,m_4,m_5)=(m_3,m_4,m_5, h-\Sigma_{i=1}^5 m_i, m_1)$,
and finally, modding by
$\caz_6$ gives rise at any {\it even} level to a $D_h^{(6)}$ partition
function:
$$\eqalign {D_h^{(6)}  & = \sum\nolimits_
{\lbrace\lambda\vert \Sigma_{i=1}^5 i m_i =3 mod 6\rbrace}\quad\vert\chi_
{\lambda}\vert ^2\quad + \sum\nolimits_{\lbrace\lambda\vert
\Sigma_{i=1}^5 i m_i =3+{k\over2} mod 6\rbrace} \chi_{\lambda}
\bar\chi_{\sigma(\lambda)}\cr
&\;+ \sum\nolimits_{\lbrace\lambda\vert \Sigma_{i=1}^5 i m_i =4k+3 mod 6
\rbrace}\chi_{\lambda} \bar\chi_{\sigma^2 (\lambda)}
+\sum\nolimits_{\lbrace\lambda\vert \Sigma_{i=1}^5 i m_i =3+{3k\over2}
mod 6\rbrace}\chi_{\lambda} \bar\chi_{\sigma^3 (\lambda)}\cr
&\; +\sum\nolimits_{\lbrace\lambda\vert \Sigma_{i=1}^5 i m_i =5k+3 mod 6
\rbrace} \chi_{\lambda} \bar\chi_{\sigma^4 (\lambda)}
+\sum\nolimits_{\lbrace\lambda\vert \Sigma_{i=1}^5 i m_i =3+{5k\over2}
mod 6\rbrace}\> \chi_{\lambda} \bar\chi_{\sigma^5 (\lambda)},\cr}
\eqno(3.92)$$
\noindent where $\sigma(m_1, m_2, m_3,m_4,m_5) =(m_2,m_3,m_4,m_5,
h-\Sigma_{i=1}^5 m_i)$.

\

\noindent {\bf E - Series}

According to the following conformal embeddings:$^{20}$
$$\widehat {su(6)}_{k=4} \subset \widehat {su(15)}_{k=1}\quad,
\quad\widehat {su(6)}_{k=6} \subset \widehat {(B_{17})}_{k=1}$$
$$\widehat {su(6)}_{k=6} \subset \widehat {(C_{10})}_{k=1}\quad,
\quad\widehat {su(6)}_{k=8} \subset \widehat {su(21)}_{k=1},\eqno(3.93)$$
\noindent we expect to find exceptional partition functions at
those levels. We will limit ourselves to only the first case in this work.

  At level $k = 4 \; (h = 10)$, we choose to start with
$D_{10}^{(2)}$. Modding by $\caz_2$ and $\caz_4$ gives $D_{10}^{(2)}$
itself, and modding by $\caz_3$, results in a combination of $D_{10}^{(2)}$
and $D_{10}^{(6)}$.
However modding by $\caz_8$ after subtracting $D_{10}^{(2)}$ with
multiplicity $12$, leads to the case III of Subsec. 2.2., which after
subtraction $D_{10}^{(2)}$ of multiplicity $12$ from the sum, we find a
modular invariant combination (having some negative integer coefficients),
which we denote by $M_{10}$:
$$\Bigr[\bigl(\sum_{\alpha=1}^8 T^\alpha+\sum_{\alpha=1}^2T^\alpha ST^2
+ST^4\bigr)SZ_1+Z_1\Bigr]=12D_{10}^{(2)}-2M_{10},\eqno(3.94)$$
\noindent where
$$\eqalignno {M_{10} & =\!\!\big\vert(\chi_{(1,1,1,1,1)}\!\!+\!\!\chi_
{(1,1,5,1,1)})
\!\!-\!\!(\chi_{(1,1,1,3,3)}\!\!+\!\!\chi_{(3,3,1,1,1)})\!\!-\!\!
(\chi_{(1,2,1,2,1)}
\!+\!\chi_{(2,1,3,1,2)})\big\vert^2\cr
& \> +\!\!\big\vert(\chi_{(1,1,1,5,1)}\!\!+\!\!\chi_{(5,1,1,1,1)})
\!\!-\!\!(\chi_{(1,1,1,1,3)}\!\!+\!\!\chi_{(1,3,3,1,1)})\!\!
-\!\!(\chi_{(1,2,1,3,1)}\!\!+\!\!\chi_{(3,1,2,1,2)})\big\vert^2\cr
&\> +\!\!\big\vert(\chi_{(1,1,1,1,5)}\!\!+\!\!\chi_{(1,5,1,1,1)})
\!\!-\!\!(\chi_{(1,1,3,3,1)}\!\!+\!\!\chi_{(3,1,1,1,1)})\!\!
-\!\!(\chi_{(1,3,1,2,1)}\!\!+\!\!\chi_{(2,1,2,1,3)})\big\vert^2\cr
& \; + 3\big\vert\chi_{(1,1,2,2,2)}+\chi_{(2,2,2,1,1)}\big\vert^2
+ 3\big\vert\chi_{(1,2,2,2,2)}+\chi_{(2,2,1,1,2)}\big\vert^2\cr
& \; + 3\big\vert\chi_{(2,1,1,2,2)}+\chi_{(2,2,2,1,1)}\big\vert^2.&(3.95)
\cr}$$
Then, modding by  $\caz_5$ gives rise to the case II, which after
subtracting $D_{10}^{(2)}$, $D_{10}^{(6)\, c.c.}$, and $M_{10}$ with
multiplicities $2/5,4/5$, and $8/5$ we find a modular invariant partition
function which we denote by $E_{10}$:
$$\biggl[\sum_{\alpha=1}^5 T^\alpha SZ_1+Z_1\biggr]=
{2\over5}\;\bigl(D_{10}^{(2)}+2D_{10}^{(6)\,c.c.}+4M_{10}+E_{10}\bigr)
,\eqno(3.96)$$
\noindent where
$$\eqalign {E_{10} & =\big\vert\chi_{(1,1,1,1,1)}\!+\!\chi_{(1,1,5,1,1)}
\!+\!\chi_{(1,2,1,2,1)}+\chi_{(2,1,3,1,2)}\big\vert^2\cr
& \>+\!\big\vert\chi_{(1,1,1,1,5)}\!+\!\chi_{(1,5,1,1,1)}
\!+\!\chi_{(1,3,1,2,1)}\!+\!\chi_{(2,1,2,1,3)}\big\vert^2\hfil\cr
& \>+\big\vert\chi_{(1,1,1,5,1)}+\chi_{(5,1,1,1,1)}
+\chi_{(1,2,1,3,1)}+\chi_{(3,1,2,1,2)}\big\vert^2\cr
& \>+\!\!\big\vert\chi_{(1,1,3,1,1)}\!\!+\!\!\chi_{(1,2,1,1,3)}
\!\!+\!\!\chi_{(1,3,2,1,2)}\big\vert^2
\!\!+\!\!\big\vert\chi_{(1,1,3,1,1)}\!\!+\!\!\chi_{(2,1,2,3,1)}
\!\!+\!\!\chi_{(3,1,1,2,1)}\big\vert^2\cr
& \>+\!\!\big\vert\chi_{(2,2,1,2,2)}\!\!+\!\!\chi_{(1,1,1,4,1)}
\!\!+\!\!\chi_{(4,1,2,1,1)}\big\vert^2
\!\!+\!\!\big\vert\chi_{(2,2,1,2,2)}\!\!+\!\!\chi_{(1,1,2,1,4)}
\!\!+\!\!\chi_{(1,4,1,1,1)}\big\vert^2\cr
& \>+\!\!\big\vert\chi_{(1,2,2,1,2)}\!\!+\!\!\chi_{(2,1,4,1,1)}
\!\!+\!\!\chi_{(1,1,1,2,1)}\big\vert^2
\!\!+\!\!\big\vert\chi_{(1,2,2,1,2)}\!\!+\!\!\chi_{(2,1,1,1,4)}
\!\!+\!\!\chi_{(1,4,1,2,1)}\big\vert^2\cr
& \>+\!\!\big\vert\chi_{(2,1,2,2,1)}\!\!+\!\!\chi_{(1,1,4,1,2)}
\!\!+\!\!\chi_{(1,2,1,1,1)}\big\vert^2
\!\!+\!\!\big\vert\chi_{(2,1,2,2,1)}\!\!+\!\!\chi_{(1,2,1,4,1)}
\!\!+\!\!\chi_{(4,1,1,1,2)}\big\vert^2\cr
& \>+\!\!\big\vert\chi_{(1,3,1,1,3)}\!\!+\!\!\chi_{(1,2,3,1,1)}
\!\!+\!\!\chi_{(1,1,2,1,2)}\big\vert^2
\!\!+\!\!\big\vert\chi_{(1,3,1,1,3)}\!\!+\!\!\chi_{(3,2,1,2,1)}
\!\!+\!\!\chi_{(2,1,1,3,2)}\big\vert^2\cr
& \>+\!\!\big\vert\chi_{(3,1,1,3,1)}\!\!+\!\!\chi_{(1,1,3,2,1)}
\!\!+\!\!\chi_{(2,1,2,1,1)}\big\vert^2
\!\!+\!\!\big\vert\chi_{(3,1,1,3,1)}\!\!+\!\!\chi_{(1,2,1,2,3)}
\!\!+\!\!\chi_{(2,3,1,1,2)}\big\vert^2.\cr}\eqno(3.97)$$
We have checked that that $E_{10}$ actually corresponds to
conformal embedding $\widehat {su(6)}_{k=4} \subset \widehat {su(15)}_{k=1}$.
However its explicit form was unknown, because of impractibility of
other methods in high rank and high level models.
 We have carried out the calculation on modding out $D_{10}^{(2)}$
 up to $\caz_{40}$, but no more exceptional partition function is found.
 Then, we strat with $D_{10}^{(6)}$, mod it out by {\it allowed} groups
 up to $\caz_{40}$, but we get the same results as above.
 For example, modding $D_{10}^{(6)}$ by $\caz_5$ gives rise to the
 following result:
$$\Bigl[\sum_{\alpha=1}^5 T^\alpha SZ_1+Z_1\Bigr]=
{2\over5}\bigl(D_{10}^{(2)\,c.c.}+2D_{10}^{(6)}+4M_{10}^{c.c.}
+E_{10}^{c.c.}\bigr),\eqno(3.98)$$
so we will not go into the details any further.

\

\

\noindent {\bf 3.6 $\widehat {g_2}$  WZW models}

\noindent As usual there exists at each level a diagonal theory
$A_h = \sum_{\lambda \in B_h}\,\vert \chi_\lambda \vert ^2$ where
$$B_h = \Bigl\lbrace \lambda\; \big\vert\; 3\leq 2m_1+m_2 < {h = k + 4}
\Bigr\rbrace,\eqno(3.99)$$
is the fundamental domain, and $\lambda= m_1\,\omega_1+m_2\,\omega_2$, but
since the centre of $G_2$ is trivial these models have no $D$ series.
However, some exceptional partition functions have been
found,$^{13,28}$ which we are going to obtain by our orbifold method.
For $\hat g_2$ theories the  factor
$(voll \, cell\, of\, Q^*/vol \, cell \, of \, Q )$ in eq. (2.1)
for the operator S is equal to $1/3$, and $\check h = 4$. The $\caz_N$
action on the characters of left$-$moving HW representations is defind due
to eq. (2.9) by
$$p \;\cdot\; \chi_{(m_1,m_2)}=e^{{2\pi i\over N}(6m_1^2+2m_2^2+6m_1m_2-14)}
\chi_{(m_1,m_2)}\eqno(3.100)$$
\noindent where $p$ is the generator of $\caz_N$.
 Thus the untwisted part of a partition function, consists of left$-$moving
 HW representations  which satisfy: $6m_1^2+2m_2^2+6m_1m_2=14\,mod\,N$.

\

\noindent {\bf E - Series}

One expects to find exceptional partition functions at levels $k=3,4$
corresponding to the the following conformal embeddings [20]:
$$\widehat {(g_2)}_{k=3} \subset \widehat {(e_6)}_{k=1}\quad,
\quad \widehat {(g_2)}_{k=4} \subset \widehat {(D_7)}_{k=1}.\eqno(3.101)$$

{\bf 1.}\  At level $k=3 \; (h=7)$, we start with $A_7$ and mod it out by
$\caz_7$ accordindg to eq. (2.7) with $N=7$; after doing the sum in
the {\it bracket}, we encounter the case I of Subsec. 2.2., which after
subtraction $A_7$ of multiplicity $2$, an exceptional partition function
 appears, which we denote by $E_7$:
$$\Bigl[\sum_{\alpha=1}^7 T^\alpha SZ_1+Z_1\Bigr]=2A_7+3E_7,\eqno(3.102)$$
\noindent where
$$E_7=\big\vert\chi_{(1,1)}+\chi_{(2,2)}\big\vert^2+2\big\vert\chi_{(1,3)}
\big\vert^2.\eqno(3.103)$$
It actually corresponds to conformal embedding
$\widehat {(g_2)}_{k=3} \subset \widehat {(e_6)}_{k=1}$.$^{13}$
 We also obtained $E_7$ in modding by $\caz_3$ but in the latter case
 we encounter the case II which after subtracting $A_7$ with
 mutiplicity $4$ we get $E_7$
$$\Bigl[\sum_{\alpha=1}^3 T^\alpha SZ_1+Z_1\Bigr]=4A_7-E_7.\eqno(3.104)$$
We have worked out all the $allowed$ moddings up to $\caz_{42}$ but no
other exceptional theories is found.

{\bf 2.}\ At level $k=4 \; (h=8)$, starting with $A_8$ and modding by
$\caz_8$ leads to the case I, which after subtraction $A_8$ of mutiplicity
$6$, an exceptional modular invariant partition function is obtained,
which we call $E_8^{(1)}$
$$\Bigr[\bigl(\sum_{\alpha=1}^8 T^\alpha+\sum_{\alpha=1}^2T^\alpha ST^2
+ST^4\bigr)SZ_1+Z_1\Bigr]=6A_8+E_8^{(1)},\eqno(3.105)$$
\noindent where
$$E_8^{(1)}=\big\vert\chi_{(1,1)}+\chi_{(1,4)}\big\vert^2+\big\vert
\chi_{(1,5)}+\chi_{(2,1)}\big\vert^2+2\big\vert\chi_{(2,2)}\big\vert^2.
\eqno(3.106)$$
It can easily be shown that $E_8$ actually corresponds to conformal embedding
$\widehat {(g_2)}_{k=4}\subset \widehat {(D_7)}_{k=1} $.$^{13}$
 In modding $A_8$ by $\caz_3$ we encounter the case II in which we can
 recognize a new exceptional partition function, which we  call
 $E_{8}^{(2)}$
$$\Bigl[\sum_{\alpha=1}^3 T^\alpha SZ_1+Z_1\Bigr]=3A_8+E_{8}^{(1)}-
E_{8}^{(2)},\eqno(3.107)$$
\noindent where
$$\eqalignno {E_{8}^{(2)}& =\big\vert\chi_{(1,1)}\big\vert^2+\big\vert
\chi_{(1,4)}\vert^2+\big\vert\chi_{(2,2)}\big\vert^2+\big\vert\chi_{(1,3)}
\big\vert^2+\big\vert\chi_{(2,3)}\big\vert^2\cr
& \> +\left(\bigl(\chi_{(1,2)}\bar\chi_{(3,1)}
+\chi_{(1,5)}\bar\chi_{(2,1)}\bigr)+c.c.\right).&(3.108)\cr}$$
This exceptional partition function, which doesn't correspond to a conformal
embedding or simple currents, was found for the first time in Ref. 28.
We have worked out all the {\it allowed} moddings up to
$\caz_{48}$ but no more exceptional partition function is found.

\

\noindent {\bf 4. Conclusions}

\noindent In this paper we have introduced an {\it orbifold-like} approach,
 as a unified method for finding all nondiagonal partition functions of a
 WZW model. For a WZW theory based on Lie group $G$, first we start with a
known theory e.g. a member of $A_h$ or $D_h$ series and divide it by
some cyclic group $\caz_N$ acting on quantum states.
In this procedure only certain group $\caz_N$'s are {\it allowed} for
a specific $\hat g_k$ theory and they lead to a modular
invariant combination. Furthermore, if all the coefficients in the
combination are positive integers, then we are asured of a physical
theory corresponding to an orbifold. What we have learned from applying
our method to
$\widehat {su(n)}$ and $\hat g_2$ in Sec. $3$ is that, the {\it allowed}
moddings are usually the ones for which $N$ is a divisor or multiple of
$m \, h$, where m is the factor
$(voll\, cell\, of\, Q/voll\, cell\, of\, Q^*)$
in the operator S in eq. (2.1), $h=k+\check h$, and
$\check h$ is the dual coxeter number of $g$.
All partition functions which may exist at a certain level are just found by
some finite set of {\it allowed} moddings, so if no partition
function appears after some finite set of moddings, one can infer that there
does not exist any partition function at that level.
With the aid of this method we have found all the known partition
functions and some new ones for $\widehat {su(n)}$ with $n=2,3,4,5,6$, and
also $\widehat {g_2}$ models.

An important feature of our method is that one can systematically  search
for exceptional partition functions, in theories with high rank groups
or/and high levels. As an example we applied our approach to
$\widehat {SU(6)}$ WZW in Subsec. 3.5., and found a new exceptional
partition function.
Work is in progress for finding exceptional partition functions of
$\hat B$, $\hat C$, $\hat D$, $\hat F_4$ models.$^{38}$
Alhough  in this paper we were concerned with WZW theories with affine
symmetry $\hat g\otimes \hat g $, i.e., with the symmetry algebras of
left$-$moving and right$-$moving sectors being the same, however our approach
is also applicable to heterotic WZW models with different algebras in their
left$-$moving and right$-$moving sectors.

Finally, we have observed that every nondiagonal theory comes from some
specific moddings, and the question arises to uderlying principle behind
these moddings. We think that in this way, it is possible to address
the basic question of classification of a WZW model.

\

\

\noindent {\bf Acknowledgements}
\noindent We would like to thank Shahin Rouhani for collaboration at
the early stage of this work. We are also grateful for very valuable
discussions with Hessam Arfaei. This work was supported in part by a research
grant from the Sharif University of Technology.

\

\

\noindent {\bf Appendix A}

\noindent As it was mentioned in Subsec. 2.2., starting with a theory
defined on a group manifold $G$ and modding by a cyclic group $\caz_N$, the
partiton function of ${G/ \caz_N}$ theory can be obtained, using the
untwisted part of partition function $Z_1$, and the generators $S$ and $T$ of
the modular group. For $N$ prime, the following equation is obtained:
$$Z(G/\caz_N)=\biggl[\sum_{\alpha=1}^N T^\alpha SZ_1+Z_1\biggr]-Z(G)
.\eqno(A.1)$$

However, It soon becomes appear that when $N$ is not prime, for generating
the full partition function (A.1), at least for each divisor $m$ of $N$,
terms in the form
$$\sum_{\alpha =1}^{\beta_m} T^\alpha ST^m SZ_1(G/\caz_N)$$
must be added into the {\it bracket} of $(A.1)$, where
$$\beta_m={\bigl(N/ [m,N]\bigr)\over\bigl[m,(N/[m,N])\bigr]}\eqno(A.2)$$
\noindent and $[\; ,\; ]$ denotes the biggest common divisor. But in
some cases, as can be seen in some of the following examples, more terms are
requierd, which have the general form of
$\sum_{\alpha =1}^{\beta_p}T^\alpha ST^p SZ_1(G/\caz_N)$, where p has some
common divisor with $N$, and is always smaller than it.
In order for a modding to lead to a modular invariant combination,
for any sum: $\sum_{\alpha =1}^{\beta_m} T^\alpha \,ST^m SZ_1 (G/\caz_N)$,
the following relation must be satisfied
$$T^{\beta_m} ST^ mSZ_1(G/\caz_N)= ST^m SZ_1 (G/\caz_N).\eqno(A.3)$$
Here we gather a list of formulas which have been used in our present
work in modding a theory by $\caz_N$'s with $N$ nonprime.
$$Z(G/\caz_4)=\biggl[\Bigl(\sum_{\alpha=1}^4 T^\alpha+ST^2\Bigr)
SZ_1+Z_1\biggr]-{1\over2}\, Z(G/\caz_2)-Z(G)\eqno(A.4)$$

\

$$\eqalign {Z(G/\caz_6)=&\!\biggr[\Bigl(\sum_{\alpha=1}^6 T^\alpha\!+\!
\sum_{\alpha=1}^3T^\alpha ST^2+\sum_{\alpha=1}^2T^\alpha ST^3\Bigr)SZ_1\!
+\!Z_1\biggr]\cr&-\! Z(G/\caz_3)\!-\! Z(G/\caz_2)\!-\!Z(G)\cr}\eqno(A.5)$$

\

$$\eqalign {Z(G/\caz_8) =&\biggr[\Bigl(\sum_{\alpha=1}^8 T^\alpha
+\sum_{\alpha=1}^2T^\alpha ST^2+ST^4\Bigr)SZ_1+Z_1\biggr]\cr
& -{1\over2}\, Z(G/\caz_4)-{1\over2}\, Z(G/\caz_2)-Z(G)\cr}\eqno(A.6)$$

\

$$Z(G/\caz_9)=\biggr[\Bigl(\sum_{\alpha=1}^9 T^\alpha+ST^3+ST^6\Bigr)
SZ_1+Z_1\biggr]-{2\over3}\, Z(G/\caz_4)-Z(G)\eqno(A.7)$$

\

$$\eqalign {Z(G/\caz_{10})\!=&\!\biggr[\Bigl(\sum_{\alpha=1}^{10} T^\alpha
+\sum_{\alpha=1}^5T^\alpha ST^2+\sum_{\alpha=1}^2T^\alpha ST^5 \Bigr)SZ_1\!
+\!Z_1\biggr]\cr&-\! Z(G/\caz_5)\!-\! Z(G/\caz_2)\!-\!Z(G)\cr}\eqno(A.8)$$

\

$$\eqalign {Z(G/\caz_{12}) =&\!\biggr[\Bigl(\sum_{\alpha=1}^{12} T^\alpha
+\sum_{\alpha=1}^3T^\alpha ST^2+\sum_{\alpha=1}^4T^\alpha ST^3
+\sum_{\alpha=1}^3ST^4+ST^6\Bigr)SZ_1\!+\!Z_1\biggr]\cr
& -\!{1\over2} Z(G/\caz_6)\!-Z(G/\caz_4)\!-\!
Z(G/\caz_3)\!-\!{1\over2}Z(G/\caz_2)\!-\!Z(G)\cr}\eqno(A.9)$$

\

$$\eqalign {Z(G/\caz_{16}) =&\!\biggr[\Bigl(\sum_{\alpha=1}^{16} T^\alpha
\!+\!\sum_{\alpha=1}^4T^\alpha ST^2\!+\!ST^4\!+\!ST^8\!+\!ST^{12}\Bigr)SZ_1\!
+\!Z_1\biggr]\cr
& -\!{1\over2} Z(G/\caz_8)\!-\!{1\over2} Z(G/\caz_4)
\!-\!{1\over2} Z(G/\caz_2)\!-\!Z(G)\cr}\eqno(A.10)$$

\

$$\eqalign {Z(G/\caz_{18}) =&\biggr[\Bigl(\sum_{\alpha=1}^{18} T^\alpha
+\sum_{\alpha=1}^9T^\alpha ST^2+ \sum_{\alpha=1}^2T^\alpha ST^3+ST^6
+\sum_{\alpha=1}^2T^\alpha ST^9\cr
&\quad+ST^{12}+\sum_{\alpha=1}^2T^\alpha ST^{15}\Bigr)SZ_1+Z_1\biggr]
-Z(G/\caz_9)-{2\over3} Z(G/\caz_{6})\cr
& -{2\over3} Z(G/\caz_{3})-Z(G/\caz_2)-Z(G)\cr}\eqno(A.11)$$

\

$$\eqalign {Z(G/\caz_{24}) =&\biggr[\Bigl(\sum_{\alpha=1}^{24} T^\alpha
+\sum_{\alpha=1}^{6}T^\alpha ST^2+ \sum_{\alpha=1}^8T^\alpha ST^3
+\sum_{\alpha=1}^{3}T^\alpha ST^4+\sum_{\alpha=1}^2T^\alpha ST^6\cr
& \quad +\sum_{\alpha=1}^{3}T^\alpha ST^8+ST^{12}\Bigr)SZ_1+Z_1
\biggr]-{1\over2} Z(G/\caz_{12})-Z(G/\caz_8)\cr
& -{1\over2} Z(G/\caz_{6})-{1\over2} Z(G/\caz_{4})-Z(G/\caz_3)
-{1\over2} Z(G/\caz_{2})-Z(G)\cr}\eqno(A.12)$$

\

$$\eqalign {Z(G/\caz_{25}) =&\biggr[\bigl(\sum_{\alpha=1}^{25} T^\alpha
+ST^5+ST^{10}+ST^{15}+ST^{20}\bigr)SZ_1+Z_1\biggr]\cr
& -{4\over5}  Z(G/\caz_{5})-Z(G)\cr}\eqno(A.13)$$

\

$$\eqalign {Z(G/\caz_{36}) = & \biggr[\Bigl(\sum_{\alpha=1}^{36} T^\alpha
+\sum_{\alpha=1}^{9}T^\alpha ST^2+\sum_{\alpha=1}^{4}T^\alpha ST^3
+\sum_{\alpha=1}^9T^\alpha ST^4+ST^6+\sum_{\alpha=1}^4T^\alpha ST^9\cr
&\quad +ST^{12}+\sum_{\alpha=1}^4T^\alpha ST^{15}+ST^{18}+ST^{24}
+ST^{30}\Bigr)SZ_1+Z_1\biggr]\cr
&-{1\over2} Z(G/\caz_{18})-{2\over3} Z(G/\caz_{12})
-Z(G/\caz_{9})-{1\over3}Z(G/\caz_{6})\cr
&-Z(G/\caz_4)-{2\over3}Z(G/\caz_{3})-Z(G)\cr}\eqno(A.14)$$

\

$$\eqalign {Z(G/\caz_{40}) =&\biggr[\Bigl(\sum_{\alpha=1}^{40} T^\alpha
+\sum_{\alpha=1}^{10}T^\alpha ST^2+ \sum_{\alpha=1}^5T^\alpha ST^4
+\sum_{\alpha=1}^8T^\alpha ST^5+ \sum_{\alpha=1}^5T^\alpha ST^8 \cr
& \quad+\sum_{\alpha=1}^{2}T^\alpha ST^{10}+ST^{20}\Bigr)SZ_1+Z_1\biggr]
-{1\over2} Z(G/\caz_{20})- Z(G/\caz_{10})\cr
& -Z(G/\caz_{8})-Z(G/\caz_5)-{1\over2}Z(G/\caz_{4})
-{1\over2} Z(G/\caz_{2})-Z(G)\cr}\eqno(A.15)$$

\

$$\eqalign{Z(G/\caz_{72})\! =\!& \biggl[ \Bigl(\sum_{\alpha=1}^{72}T^{\alpha}
\!+\!\sum_{\alpha=1}^{18}T^{\alpha}ST^2\!+\!\sum_{\alpha=1}^8 T^{\alpha}ST^3
\!+\!\sum_{\alpha=1}^9 T^{\alpha}ST^4\!+\!\sum_{\alpha=1}^2 T^{\alpha}ST^6\cr
& \quad+\!\sum_{\alpha=1}^9 T^{\alpha}ST^8
\!+\!\sum_{\alpha=1}^8 T^{\alpha}ST^9+ST^{12}
\!+\!\sum_{\alpha=1}^8 T^{\alpha}ST^{15}\!+\!\sum_{\alpha=1}^2 T^{\alpha}
ST^{18}\cr
& \quad+\!ST^{24}\!+\!\sum_{\alpha=1}^2 T^{\alpha}ST^{30}\!+\!ST^{36}\!
+\!ST^{48}\!+\!ST^{60}\Bigr)SZ_1\!+\!Z_1\biggr]\cr
& -{1\over2} Z(G/\caz_{36})-{2\over3} Z(G/\caz_{24})
-{1\over2}Z(G/\caz_{18})-{1\over3}Z(G/\caz_{12})\cr
& -Z(G/\caz_{9})-Z(G/\caz_{8})-{1\over3}Z(G/\caz_{6})
-{1\over2}Z(G/\caz_{4})\cr
& -{2\over3}Z(G/\caz_{3})-{1\over2}Z(G/\caz_{2})-Z(G)\cr}\eqno(A.16)$$

\

\noindent {\bf References:}
\item  {1.} A. M. Polyakov, {\it Sov. Phys. JETP. Lett.} {\bf 12}, 381 (1970)
\item  {2.} A. A. Belavin, A. M. Polyakov, A. B. Zamolodchikov, {\it Nucl.
Phys.} {\bf B241}, 333 (1984)
\item  {3.} D. Friedan, {\it Notes on String Theory and Two Dimensional
Conformal Field Theory, in Unified String Theories}, eds. M. Green and D.
Gross (World Scientific, Singapore, 1986)
\item  {4.} W. Nahm, {\it'Conformally Invariant Quantum Field Theories in Two
Dimensions'}, (World Scientific, Singapore, 1992)
\item  {5.} E. Witten, {\it Commun. Math. Phys.} {\bf 92}, 455 (1984)
\item  {6.} V. K. Knizhnik and A. B. Zamolodchikov, {\it Nucl. Phys.}
{\bf B247}, 83 (1984)
\item  {7.} P. Goddard, A. Kent, D. Olive, {\it Commun. Math. Phys.}
{\bf 103},105 (1986)
\item  {8.} I. Affleck, {\it Nucl. Phys.} {\bf B265 [FS15]}, 409 (1986)
\item  {9.} D. Gepner and E. Witten, {\it Nucl. Phys.} {\bf B278}, 493 (1986)
\item  {10.} A. Cappeli, C. Itzykson, J. B. Zuber, {\it Nucl. Phys.}
{\bf B280}, 445 (1987); {\it Commun. Math. Phys.} {\bf 113}, 1 (1987)
\item {11.} T. Gannon, {\it 'The Classification of Affine SU(3) Modular
Invariant Partition Functions', Carleton preprint, hep-th 9209042,} (1992)
\item {12.} M. Bauer and C. Itzykson, {\it Commun. Math. Phys.}
{\bf 127}, 617 (1990)
\item {13.} P. Chirste and F. Ravanini, {\it Int. J. Mod. Phys.} {\bf A4},
 897 (1988) 897
\item {14.} C. Itzykson, {\it Nucl. Phys. (Proc. Suppl.)} {\bf 5B}, 150
(1988);T. Gannon, {\it 'WZW commutants, Lattices, and Level 1 Partition
Functions',\ Carleton preprint, hep-th 9209043}, (1992)
\item {15.} F. Ardalan, H. Arfaei, {\it Sharif University preprint}
 (Unpublished)
 and {\it Proceeding of the III Regional Conference Islamabad} (World
 Scientific, 1989)
\item {16.} D. Bernard, {\it Nucl. Phys.} {\bf B288}, 628 (1987)
\item {17.} G. Felder, K. Gawedzki, A. Kupianen, {\it Commun. Math. Phys.}
{\bf 117}, 127 (1988)
\item {18.} A. N. Schellekens, S. Yankielowicz, {\it Nucl. Phys.} {\bf B327},
 673 (1989); {\it Phys. Lett} {\bf B227}, 387 (1989)
\item {19.} A. N. Schellekens and S. Yankielowicz, {\it Nucl. Phys.}
{\bf B334}, 67 (1990)
\item {20.} F. Bais and P. Bowknegt, {\it Nucl. Phys.} {\bf B279}, 561 (1987)
\item {21.} A. N. Schellekens and N. B. Warner, {\it Phys. Rev.} {\bf D34},
 3092 (1986)
\item {22.} P. G. Bouwknegt and W. Nahm, {\it Phys. Lett.} {\bf 184 B}, 359
 (1987)
\item {23.} R. Dijkgraaf and E. Verlinde, {\it in Proceeding of the Annecy
Conference on Conformal Field Theory} (March 1988)
\item {24.} G. Moore and N. Seiberg, {\it Nucl. Phys.} {\bf B313}, 16 (1989)
\item {25.} A. Font, {\it Mod. Phys. Lett.} {\bf A6}, 3265 (1991)
\item {26.} N. P. Warner, {\it Commun. Math. Phys.} {\bf 127}, 71 (1990)
\item \ \ \ P. Robertson, and H. Terao,{\it Int. J. Mod. Phys.}{\bf A7},
2207 (1992)
\item {27.} A. Shirzad and H. Arfaei, {\it Modular Invariant Partition
Functions and Method of Shift Vectors, BONN-HE-93-1, SUTDP/93/71/1}, (1993)
\item {28.} D. Verstegen, {\it Nucl. Phys.} {\bf B346}, 349 (1990)
\item {29.} F. Ardalan, H. Arfaei, S. Rouhani, {\it Int. J. Mod. Phys.}
{\bf A}, 4763 (1991)
\item {30.} V. G. Kac, {\it Infinite Dimensional Lie Algebras}
(Cambridge University Press, 1985)
\item {31.} P. Goddard and D. Olive, {\it Int. J. Mod. phys.} {\bf A1}, 303
 (1986)
\item {32.} D. Gepner, {\it Nucl. Phys.} {\bf B287}, 111 (1987)
\item {33.} L. Dixon, J. Harvay, C. vafa, E. Witten, {\it Nucl. Phys.}
{\bf B261}, 678 (1985); {\it Nucl. Phys.} {\bf B274}, 285 (1986)
\item {34.} R. Dijkgraaf, C. Vafa, E. Verlinde, H. Verlinde
{\it Commun. Math. Phys.} {\bf 123},458 (1989)
\item {35.} C. Ahn, M. A. Walton, {\it Phys. Lett.} {\bf B223}, 343 (1989)
\item {36.} L. Dixon, D. Friedan, E. Martinec, S. Schenker, {\it Nucl. Phys.}
{\bf B282}, 13 (1987)
\item {37.} S. Hamidi, C. Vafa, {\it Nucl. Phys.} {\bf B279}, 465 (1987)
\item {38.} M. R. Abolhassani, F. Ardalan (In prepration)

\end